\documentclass[runningheads]{llncs}

\newcommand{\iffextendedver}[2]{#1}
\newcommand{\ifextendedver}[1]{\iffextendedver{#1}{}}
\newcommand{\ifconfver}[1]{\iffextendedver{}{#1}}
\newcommand{\iffwithappendix}[2]{\iffextendedver{#1}{#2}}
\newcommand{\ifwithappendix}[1]{\iffextendedver{#1}{}}

\usepackage{xspace}
\usepackage{microtype}
\usepackage[svgnames]{xcolor}
\usepackage{graphicx}
\usepackage{amsmath,amssymb,amscd,stmaryrd,url,mathrsfs}
\usepackage{calligra}
\usepackage{array}
\usepackage{paralist}
\usepackage{subfig}
\usepackage{booktabs}
\usepackage{colortbl,multicol,multirow}
\usepackage{varwidth}
\usepackage[nocompress]{cite}

\usepackage[noend,ruled,vlined]{algorithm2e}

\SetCommentSty{mycommfont}
\usepackage{wrapfig}
\usepackage{orcidlink}

\usepackage{enumerate}
\usepackage{enumitem}
\usepackage[disable]{todonotes}

\usepackage{tikz}
\usetikzlibrary{shapes,arrows,automata,positioning,calc}
\usetikzlibrary{fit,backgrounds,decorations.pathreplacing}

\usepackage{hyperref}
\hypersetup{
    colorlinks,
    linkcolor={red!50!black},
    citecolor={cyan!50!black},
    urlcolor={blue!80!black}
}

\gappto{\UrlBreaks}{\UrlOrds}
\usepackage[capitalize]{cleveref} %

\newcommand{\beginparagraph}[1]{\medskip\noindent\textbf{#1}}
\newcommand{\littleparagraph}[1]{\smallskip\noindent\textit{#1}}

\tikzstyle{con}=[align=center]
\tikzstyle{cona}=[con,left]
\tikzstyle{conb}=[con,right]
\tikzstyle{state}=[circle,draw,minimum size=14pt,inner sep=0pt]
\tikzstyle{astate}=[state,double]
\tikzstyle{ustate}=[state,white,fill=black,line width=1mm]
\tikzstyle{smallstate}=[circle,draw,minimum size=10pt,inner sep=0pt]
\tikzstyle{asmallstate}=[smallstate,double]
\tikzstyle{dtnode}=[ellipse,draw,minimum size=14pt,inner sep=0pt]
\tikzstyle{dtleaf}=[rectangle,draw,minimum size=14pt,inner sep=0pt]

\newcommand{\FORGET}[1]{}

\newcommand{\ShowComments}[2]{#2}    %

\newcommand{\bjcom}[1]{\ShowComments{{\color{blue} {BJ: #1}}}{}}

\newcommand{\longversion}[1]{}

\def\squareforqed{\hbox{\rlap{$\sqcap$}$\sqcup$}}
\def\qed{\ifmmode\squareforqed\else{\unskip\nobreak\hfil
\penalty50\hskip1em\null\nobreak\hfil\squareforqed
\parfillskip=0pt\finalhyphendemerits=0\endgraf}\fi}

\newcommand{\prefixes}{{U}}
\newcommand{\prefixmap}{\mathcal{U}}

\newcommand{\domof}[1]{Dom(#1)}

\newcommand{\tuple}[1]{\langle #1\rangle}
\newcommand{\set}[1]{\lbrace #1\rbrace}
\newcommand{\vect}[2]{#1_1, \ldots, #1_{#2}}
\newcommand{\setcomp}[2]{\set{#1 ~:~ #2}}

\newcommand{\with}{~|~}
\newcommand{\sizeof}[1]{|#1]}

\newcommand{\domain}{\mathcal{D}}
\newcommand{\binrelations}{\mathcal{R}}

\newcommand{\act}{\alpha}

\newcommand{\acts}{\Sigma}

\newcommand{\emptyword}{\epsilon}
\newcommand{\dword}{w}
\newcommand{\ddword}{w'}
\newcommand{\actsof}[1]{Acts(#1)}
\newcommand{\valsof}[1]{Vals(#1)}
\newcommand{\Lang}{\mathcal{L}}

\newcommand{\langstack}{\Lang_{\mathit{Stack}}}

\newcommand{\dact}[2]{{#1(#2)}}

\newcommand{\Langof}[1]{\Lang(#1)}

\newcommand{\getmemorable}[3]{\mathit{mem}_{#3}(#1)}

\newcommand{\guardsof}[3]{{\cal G}_{#2}(#1,#3)}

\newcommand{\rootof}[1]{\mathit{root}(#1)}

\newcommand{\ctnode}{N}
\newcommand{\leafnode}{N}
\newcommand{\rapof}[1]{\mathit{rp}(#1)}
\newcommand{\lcaof}[2]{\mathit{lca}(#1,~ #2)}
\newcommand{\Siftfn}{\mathit{Sift}}
\newcommand{\Refinefn}{\mathit{Refine}}
\newcommand{\Expandfn}{\mathit{Expand}}
\newcommand{\Analyzefn}{\mathit{Analyze}}
\newcommand{\auto}{\mathcal{A}}
\newcommand{\locs}{L}
\newcommand{\loc}{l}

\newcommand{\initloc}{\loc_0}

\newcommand{\vars}{\mathcal{X}}

\newcommand{\transitions}{\Gamma}
\newcommand{\varsof}[1]{\vars(#1)}

\newcommand{\remap}{\pi}

\newcommand{\valuation}{\mu}

\newcommand{\upath}{\tau}

\newcommand{\uvtree}{T}
\newcommand{\pathcondof}[1]{{\cal G}_{#1}}

\newcommand{\freshof}[1]{\mathit{fresh}(#1)}
\newcommand{\sdtfunction}[2]{{\cal L}[#1,#2]}

\newcommand{\ancestorsof}[1]{\mathcal{V}(#1)}
\newcommand{\symbsuffof}[1]{\mathit{suff}(#1)}

\newcommand{\symbsuff}{\vec{v}}
\newcommand{\csuff}{\symbsuff}
\newcommand{\symbalpha}{\vec{\alpha}}
\newcommand{\symbsuffs}{\mathcal{V}}

\newcommand{\dval}{\mathtt{d}}
\newcommand{\repr}[2]{\mathtt{d}_{#1}^{#2}}

\newcommand{\plainequiv}[2]{\equiv_{#2}}
\newcommand{\remapequiv}[3]{\simeq_{#2}^{#3}}

\newcommand{\lengthof}[1]{|#1|}

\newcommand{\dvaloverbar}{\overline{\dval}}
\newcommand{\dvalprimeoverbar}{\overline{\dval'}}
\newcommand{\xoverbar}{\overline{x}}
\newcommand{\poverbar}{\overline{p}}

\newcommand{\hypo}{\mathcal{H}}

\newcommand{\newtreeoracle}[1]{{\cal O}}
\newcommand{\treeoracle}{{\cal T}}

\newcommand{\strans}[3]{ \frac{#1 \with #2}{#3}}
\newcommand{\true}{{\it true}}

\newcommand{\spset}[1]{U}
\newcommand{\asset}[1]{As}
\newcommand{\epset}[1]{U^{+}}

\newcommand{\eqleaf}[1]{=_{DT}}
\newcommand{\neqleaf}[1]{\neq_{DT}}

\newcommand{\treequeryhyp}[2]{\ensuremath{ \hypo({#1},{#2}) }}
\newcommand{\treequerysul}[2]{\ensuremath{ \Lang({#1},{#2}) }}

\newcommand{\invU}{\prefixmap^{\scalebox{0.3}[1.0]{\( - \)}1}}

\newcommand{\eg}{e.g.\xspace}
\newcommand{\ie}{i.e.\xspace}

\newsavebox\epsr
\newsavebox\epsl
\newsavebox\popr
\newsavebox\popl
\newsavebox\poppopl
\newsavebox\poppopr

\newsavebox\xleft
\newsavebox\xmid
\newsavebox\xright

\newsavebox\sbtrueacc
\newsavebox\sbtruerej
\newsavebox\sbtruetrueacc
\newsavebox\sbtruetruerej
\newsavebox\sbeq
\newsavebox\sbeqeq

\tikzstyle{innerdt} = [draw,rectangle,inner sep=5pt]
\tikzstyle{leafdt} = [draw,rectangle,inner sep=5pt,align=left]
\tikzstyle{sdtloc} = [draw,circle,inner sep=2pt]
\tikzstyle{sdtfloc} = [draw,double,circle,inner sep=2pt]
\tikzstyle{edgegraph} = [shorten >=1pt,->,font=\small,scale=.5,every node/.style={scale=.5}]

\newcommand{\push}{\mathsf{push}}
\newcommand{\pop}{\mathsf{pop}}

\newcommand{\SL}[1]{\ensuremath{{SL}^{#1}}\xspace}
\newcommand{\SLstar}{\SL{*}}
\newcommand{\SLlambda}{\SL{\lambda}}
\newcommand{\SLCT}{\SL{CT}}

\ifconfver{
\setlength{\floatsep}{8pt plus 1pt minus 2pt}
\setlength{\textfloatsep}{8pt plus 1pt minus 2pt}
}

\title{Scalable Tree-based Register Automata Learning
  \ifextendedver{(Extended Version with Appendices)\thanks{A version of this
      paper without appendices appears in the proceedings of TACAS~2024.}}}
\titlerunning{Scalable Tree-based Register Automata Learning}
\author{Simon Dierl\inst{1}\protect\orcidlink{0000-0001-9730-9335}
   \and Paul Fiterau-Brostean\inst{2}\protect\orcidlink{0000-0002-5185-0035}
   \and Falk Howar\inst{1}\protect\orcidlink{0000-0002-9524-4459}
   \and Bengt Jonsson\inst{2}\protect\orcidlink{0000-0001-7897-601X}
   \and Konstantinos Sagonas\inst{2,3}\protect\orcidlink{0000-0001-9657-0179}
   \and Fredrik T{\aa}quist\inst{2}\protect\orcidlink{0000-0003-4066-9078}}
\institute{Technical University of Dortmund, Dortmund, Germany
      \and Uppsala University, Uppsala, Sweden
      \and National Technical University of Athens, Athens, Greece}
\authorrunning{S. Dierl et al.}

\begin{document}

\maketitle

\begin{abstract}
Existing active automata learning (AAL) algorithms have demonstrated their
potential in capturing the behavior of complex systems (e.g., in analyzing
network protocol implementations).
The most widely used AAL algorithms generate finite state machine models,
such as Mealy machines.
For many analysis tasks, however, it is crucial to generate richer classes
of models that also show how relations between data parameters affect system
behavior. Such models have shown potential to uncover critical bugs, but their
learning algorithms do not scale beyond small and well curated experiments.
In this paper, we present \SLlambda, an effective and scalable register automata
(RA) learning algorithm that significantly reduces the number of tests required
for inferring models. It achieves this by combining a tree-based cost-efficient
data structure with mechanisms for computing short and restricted tests.
We have implemented \SLlambda as a new algorithm in RALib. We evaluate its
performance by comparing it against \SLstar, the current state-of-the-art RA
learning algorithm, in a series of experiments, and show superior performance
and substantial asymptotic improvements in bigger systems.

\end{abstract}

\section{Introduction}
\label{sec:introduction}

\emph{Model Learning} (aka \emph{Active Automata Learning} (AAL)~\cite{Angluin:regset,RiSh:inference,Vaandrager:cacm}) infers automata models that represent
the dynamic behavior of a software or hardware component from tests.
Models obtained through (active) learning
have proven useful for many purposes, such as
analyzing security
protocols~\cite{ShLe:icdcs07,Groz:isola12,Ruiter2015ProtocolStateFuzzing,DTLS@USENIX-20,Automata@NDSS-23},
mining APIs~\cite{ABL:mining}, 
supporting model-based testing~\cite{HHNS:modelgeneration,Walkinshaw:ictss10,Tappler@ICST-2017}
and conformance testing~\cite{AKTVW:brp}.
The AAL algorithms employed in these works are efficient and supported by
various domain-specific optimizations (e.g.,~\cite{HungarNS03}), but they
all generate finite state machine (FSM) models, such as Mealy machines.

For many analysis tasks, however, it is crucial for models to also be able to
describe \emph{data flow}, i.e., constraints on data parameters that are
passed when the component interacts with its environment, as well as the
mutual influence between dynamic behavior and data flow. For instance, models
of protocol components must describe how different parameter values in
sequence numbers, identifiers, etc. influence the behavior, and
vice versa. 
Existing techniques for extending AAL to \emph{Extended FSM} (EFSM)
models~\cite{CHJS:faoc-16,AJU14,BHLM:ras} take several different approaches.
Some reduce the problem to inferring FSMs by using manually supplied 
abstractions on the data domain~\cite{AJU14}, which requires insight
into the control/data dependencies of a system under learning (SUL). 
Others extend AAL for finite state models by allowing transitions to
contain predicates over rich data domains, but cannot generate state 
variables to model data dependencies between consecutive 
interactions~\cite{Drews2017LearningSymbolicAutomata,Maler14}.
Finally, there exist extensions of AAL to EFSM models with guards and 
state variables, such as \emph{register automata}~\cite{Aarts2012AutomataLearningCounterexample,Aarts2015LearningRegisterAutomata,CHJS:faoc-16}.
While their potential has been shown by being able to uncover critical bugs
in e.g., TCP implementations~\cite{FiHo:fmics17,Prognosis@SIGCOMM-21}, their
learning algorithms do not scale beyond small and well curated experiments.

We follow the third line of works and address the scalability of register
automata (RA) learning algorithms in our work. The main challenge when scaling
AAL algorithms is reducing the number of tests that learners perform on a SUL.
Generally, these tests are sequences of actions
of the form $u \cdot v$, where $u$ is the prefix and $v$ the suffix
of the sequence. 
Tests $u\cdot v$ and $u'\cdot v$ are then used to determine if prefixes 
$u$ and $u'$ can be distinguished based on the SUL's output triggered by $v$.
When inferring RA models, prefixes are sequences of actions with data values,
e.g., $\push(1)\,\push(2)$, and suffixes are sequences of actions with symbolic
parameters, e.g., $\pop(p_1)\,\pop(p_2)$, that, when instantiated, can incur a number of tests that
is exponential in the length of the suffix for identifying dependencies
between prefix values and suffix parameters, e.g., different test outcomes for
$(p_1=2 \land p_2=1)$ and $(p_1=2 \land p_2=3)$, and for distinguishing
prefixes based on suffixes.
To make register automata learning scalable, it is crucial to reduce the use
of suffixes in tests along three dimensions:
\begin{inparaenum}[(i)]
\item First, it is important to use only \emph{few tests}.
\item Second, when using suffixes in tests, \emph{shorter suffixes}
  should be preferred over longer ones.
\item Third, it is essential to \emph{restrict tests to relevant
  dependencies} between prefix values and suffix parameters instead of
  bluntly testing all possible dependencies.
\end{inparaenum}

In this paper, we present the \SLlambda algorithm for learning register
automata which achieves scalability by optimizing the use of tests and
suffixes in tests in the three stated dimensions. \SLlambda uses a
\emph{classification tree} as a data structure, constructs a minimal
prefix-closed set of prefixes and a suffix-closed set of \emph{short} and
\emph{restricted} suffixes for identifying and distinguishing locations,
transitions, and registers.
Technically, we adopt the idea of using a classification tree from learners
for FSMs~\cite{KeVa:book,IHS2014} where it proved very successful for reducing
tests. We also adopt the technique of computing short suffixes incrementally
in order to keep them short~\cite{Angluin:regset,IHS2014}.
This has not been studied for RAs before and leads to an improved worst case
complexity compared to state-of-the-art approaches (\cref{thm:complexity}).
Finally, we show how suffixes can be restricted to relevant data dependencies,
which is essential for achieving scalability (\cref{sec:algorithm}).

We have implemented \SLlambda as a new algorithm in the
RALib\footnote{RALib is available at \url{https://github.com/LearnLib/ralib}.}
tool~\cite{Cassel2015RalibLearnlibExtension}.
For comparison, we have also implemented in RALib the \SLCT algorithm
that uses a classification tree but relies on suffixes from counterexamples
instead of computing short suffixes from inconsistencies.
We evaluate the \SLlambda algorithm by comparing its performance
against the \SLstar~\cite{CHJS:faoc-16} and \SLCT algorithms in a series of
experiments, confirming that:
\begin{inparaenum}[(i)]
\item classification trees scale much better than observation tables for
  register automata,
\item using restricted suffixes leads to a dramatic reduction of tests for
  all compared algorithms, and
\item computing short suffixes from inconsistencies outperforms using suffixes
  from counterexamples.
\end{inparaenum}

\medskip
\noindent
\textbf{Related Work.}
For a broad overview of AAL refer to the survey paper of de la Higuera~\cite{Higuera2005BibliographicalStudyGrammatical} from 2005 and to a more recent paper by Howar and Steffen~\cite{Howar2018ActiveAutomataLearning}.

Learning beyond DFAs has been investigated for many models aside from register automata. For example, algorithms have been presented for workflow Petri nets~\cite{Esparza2011LearningWorkflowPetri}, data automata~\cite{Garg2013LearningUniversallyQuantified}, generic nondeterministic transition systems~\cite{Volpato2014ActiveLearningNondeterministic}, symbolic automata~\cite{Drews2017LearningSymbolicAutomata}, one-timer automata~\cite{Vaandrager2021LearningMealyMachines}, and systems of procedural automata~\cite{Frohme2021CompositionalLearningMutually}.
Learning of register automata has been performed by combining a FSM learner with the Tomte front-end~\cite{Aarts2012AutomataLearningCounterexample,Aarts2015LearningRegisterAutomata}. A different approach using bespoke RA learning algorithms~\cite{Howar2012InferringCanonicalRegister,Merten2012DemonstratingLearningRegister} has been implemented in RALib.
Active learning algorithms for nominal automata,
which extend FSMs to infinite alphabets and infinite sets of states,
have also been developed~\cite{Moerman2017LearningNominal}.
While the expressivity of nominal DFAs is equivalent to that
of deterministic register automata with equality,
nominal automata do not represent registers
symbolically but through permutations on infinite sets,
leading to big models (e.g., for storing some data value twice)
and active learning algorithms with a high query complexity.

Applications of AAL are diverse. Active learning enables the generation of behavioral models for software~\cite{Sun2015TlvAbstractionTesting,Schuts2016RefactoringLegacySoftware}, e.g. for network protocol implementations~\cite{Ruiter2015ProtocolStateFuzzing,}, enabling security analyses and model checking~\cite{Aarts2015GeneratingModelsInfinite,FiterauBrostean2016CombiningModelLearning,FiterauBrostean2017ModelLearningModel}. It can be used in testing~\cite{Margaria2004EfficientTestBased,Shahbaz2014AnalysisTestingBlack} and to enable formal analyses~\cite{Vaandrager:cacm}. Finally, it can be combined with passive learning approaches to support life-long learning~\cite{Frohme2021NeverStopContext}.
More theoretical advances include the use of Galois connections to model SUL-oracle mappers~\cite{Linard2019LearningUnionsK} and the introduction of apartness~\cite{Vaandrager2022NewApproachActive}, to formalize state~distinction.

\medskip
\noindent
\textbf{Outline.}
We present the key ideas in tree-based learning of RA informally in the next
section, before providing formal definitions of basic concepts in
\cref{sec:dlra}.
\Cref{sec:algorithm,sec:evaluation,sec:correctness} present the \SLlambda
algorithm, its properties, and the experimental evaluation of its performance.
The paper ends with few concluding remarks.

\section{Main Ideas}
\label{sec:ideas}

\begin{figure}[t]
  \centering
  \begin{tikzpicture}[shorten >=1pt,->,font=\small]
	\tikzstyle{none}=[circle,minimum size=14pt,inner sep=0pt]
	\tikzstyle{final}=[circle, draw,style=double,minimum size=14pt,inner sep=0pt]
	\tikzstyle{vertex}=[circle,draw,style=double,minimum size=17pt,inner sep=0pt]
	
	\node[none]   (nn) at (0.2,0)  {};
	\node[vertex] (l0) at (1,0)     {$l_0$};
	\node[vertex] (l1) at (5,0)   {$l_1$};
	\node[vertex] (l2) at (9,0)     {$l_2$};   
	
	\node[above] (l0r) at (l0.north) {$\scriptstyle\emptyset$};
	\node[above] (l1r) at (l1.north) {$\scriptstyle\{x_1\}$};
	\node[above] (l2r) at (l2.north) {$\scriptstyle\{x_1, x_2\}$};
	
	\draw (nn) -- (l0);
	
	\path (l0) edge[bend left=10] node[above] {
		$\strans{\push(p)}{true}{x_1:=p}$} (l1);
	
	\path (l1) edge[bend left=10] node[below] {
		$\strans{\pop(p)}{p=x_1}{-}$} (l0);
	
	\path (l1) edge[bend left=10] node[above] {
		$\strans{\push(p)}{true}{x_1:=x_1,x_2:=p}$} (l2);
	
	\path (l2) edge[bend left=10] node[below] {
		$\strans{\pop(p)}{p=x_2}{x_1:=x_1}$} (l1);
	
\end{tikzpicture}
  \caption{Register automaton accepting language of stack with capacity two.}
  \label{fig:demo:stack}
\end{figure}
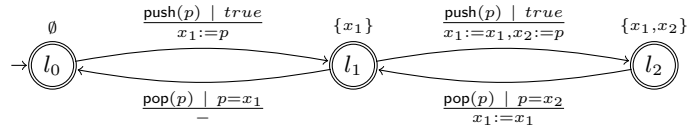

In this section, we introduce the main ideas behind the \SLlambda algorithm.
As illustrating example, we will use a stack of capacity two, which stores a sequence of natural numbers.
The stack supports the operations $\push$ and $\pop$, both of which take one natural number as a parameter.
The operation $\push(\dval)$ succeeds if the stack is not full, i.e., contains at most one element;
the operation $\pop(\dval)$ succeeds if the last pushed and not yet popped element is $\dval$.
Let a \emph{symbol} denote an operation with data value, such as $\push(1)$,
and let $\langstack$ denote the prefix-closed language consisting of the words of symbols representing sequences of successful operations.
\Cref{fig:demo:stack} shows an acceptor for $\langstack$.
The initial location $l_0$ corresponds to an empty stack, location $l_1$ corresponds to a stack with one element, and $l_2$ to a location where the stack is full.
There is also an implicit sink location for each word that is not accepted by $\langstack$, \eg pushing a third element, or popping a non-top element.
In each location, registers contain the elements in the stack: for $i= 0,1,2$, location $l_i$ has $i$ registers, where the register with the
highest index contains the topmost stack element.

The task of the \SLlambda algorithm is to learn the acceptor in \cref{fig:demo:stack}
in a black-box scenario, i.e., knowing only the operations ($\push$ and $\pop$) and the relations that may
be used in guards (here tests for equality), by asking two kinds of queries.
A \emph{membership query} asks whether a word $w$ is in $\Lang$; it can be realized by a simple test.
An \emph{equivalence query} asks whether a hypothesis RA accepts $\Lang$; if so, the query is answered by $yes$,
  otherwise by a \emph{counterexample}, which is a word on which the hypothesis and $\Lang$ disagree;
  in a black box setting it is typically approximated by a conformance testing algorithm.
Like other AAL algorithms, \SLlambda iterates a cycle in which membership queries are used to construct a hypothesis, which
is then subject to validation by an equivalence query. If a counterexample is found, hypothesis construction is resumed, etc., until
a hypothesis agrees with $\Lang$.

Classical AAL algorithms that learn DFAs maintain
an expanding set of words, $S_p$, called \emph{short prefixes}, and
an expanding set of words, called \emph{suffixes}, which induce an equivalence relation $\equiv$ on prefixes, defined by
$u \equiv u'$ iff $uv \in \Lang \Leftrightarrow u'v \in \Lang$ for all suffixes $v$;
this allows equivalence classes of prefixes to represent states in a DFA.
The \SLlambda algorithm maintains a set $\prefixes$ of data words called
\emph{prefixes}, which is the union of $S_p$ and one-symbol extensions of
elements in $S_p$.
Instead of suffixes,
$\SLlambda$ maintains a set $\symbsuffs$ of \emph{symbolic suffixes}, each of which is a \emph{parameterized} word, i.e., a word where data values are replaced by parameters $\vect pm$.
For each prefix $u$, say $\push(0)$, and symbolic suffix~$\symbsuff$, say $\push(p_1)\pop(p_2)$, membership in $\Lang$ of words of form $u\symbsuff$
depends on the relation between the data values of $u$ and the parameters $p_1,p_2$ of $\symbsuff$, which
in $\SLlambda$ is represented by a function $\sdtfunction{u}{\symbsuff}$ with parameters $x_1$ (representing the data value of $u$), $p_1$, and $p_2$.
In this case $\sdtfunction{u}{\symbsuff}(x_1,p_1,p_2)$ is $+$ iff $p_2 = p_1$ and $-$ otherwise.
In $\SLlambda$, such functions are represented as decision trees of a specific form.
\begin{wrapfigure}[7]{r}{0.3\textwidth}
  \ifconfver{\vspace*{-1em}}
  \centering
  \begin{tikzpicture}[edgegraph]
  \node[ustate] at (0,0) (l0) {};
  \node[ustate] at (2.5,0) (l1) {};
  \node[astate] at (5.5,0.9) (l2) {};
  \node[state]  at (5.5,-0.9) (l3) {};
  
  \path (l0) edge[] node [above] {$\true$} (l1);
  \path (l1) edge[] node [above] {$p_2 = p_1$} (l2);
  \path (l1) edge[] node [below] {$p_2 \neq p_1$} (l3);
\end{tikzpicture}
  \caption{Decision tree for $\sdtfunction{u}{\symbsuff}(x_1,p_1,p_2)$.}
  \label{fig:sdt:push-pop}
\end{wrapfigure}
\Cref{fig:sdt:push-pop} shows the decision tree for the just described function.
Note that it checks constraints for parameters one at a time: first the
constraint on only $p_1$ (which is $\true$), and thereafter the constraint on $p_2$ (a comparison with $p_1$).
Two prefixes $u$, $u'$ are then equivalent w.r.t.\ $\symbsuffs$ if $\sdtfunction{u}{\symbsuff}$ and $\sdtfunction{u'}{\symbsuff}$ are ``isomorphic modulo renaming''
for all $\symbsuff \in \symbsuffs$ (details in \cref{sec:algorithm}).

Functions of form $\sdtfunction{u}{\symbsuff}$ are generated by so-called \emph{tree queries}, which perform  membership queries for relevant combinations of relations between
data values in $u$ and parameters in $\symbsuff$, and summarize the results in a canonical way.
The tree query above requires five membership queries. $\SLlambda$ employs techniques for reducing this number by restricting the symbolic suffix; see end of this section.

Initially, $S_p$ and $\symbsuffs$ contain only the empty sequence $\epsilon$.
Since $\epsilon$ is a short prefix, one-symbol extensions, $\push(0)$ and $\pop(0)$,
are entered into $\prefixes$. Tree queries are performed for the prefixes in $\prefixes$ and the empty suffix, revealing that $\push(0)$ is accepted and $\pop(0)$ is rejected. Thus,
$\push(0)$ cannot be distinguished from $\epsilon$, but $\pop(0)$ can, so it must lead to a new location, hereafter referred to as the \emph{sink}, which is therefore added to $S_p$.
One-symbol extensions of $\pop(0)$, in this case $\pop(0)\push(1)$ and $\pop(0)\pop(1)$, are added to $\prefixes$ and tree queries for them and the empty suffix are performed,
revealing that they cannot be separated from the sink.
At this point, we can formulate hypothesis $\hypo_0$ in \cref{fig:demo:stack:01}(left)
from $S_p$, $\prefixes,$ and the computed decision trees.

\begin{figure}[t]
  \centering
  \setlength\fboxsep{-2pt}
  \begin{varwidth}{.3\textwidth}
    \centering
    \fbox{\begin{tikzpicture}[shorten >=1pt,->,font=\small]
  \tikzstyle{none}=[circle,minimum size=14pt,inner sep=0pt]
  \tikzstyle{final}=[circle, draw,style=double,minimum size=14pt,inner sep=0pt]
  \tikzstyle{vertex}=[circle,draw,minimum size=15pt,inner sep=0pt]
  
  \node[none]   (nn) at (0,0)  {};
  \node[final]  (l0) at (0,-0.8)    {$l_0$};
  \node[vertex] (ls) at (0,-2.3)    {\tiny sink};
  \node[left]  (l0r) at (l0.west)   {$\scriptstyle\emptyset$};
  \node[left]  (lsr) at (ls.west)   {$\scriptstyle\emptyset$};
  \node[none]   (H)  at (-0.7,-1.4) {\textcolor{blue}{\ \ $\hypo_0$\ }};
 
  \draw (nn) -- (l0);
  
  \path (l0) edge[loop right] node[right] {
    $\strans{\push(p)}{true}{-}$} (l0);
  \path (l0) edge[] node[right] {
    $\strans{\pop(p)}{true}{-}$} (ls);
  \path (ls) edge[loop right] node[right,align=center] {
    $\strans{\push(p)}{true}{-}$ \\
    $\strans{\pop(p)}{true}{-}$} (ls);
\end{tikzpicture}}
  \end{varwidth}
  \setlength\fboxsep{-.5em}
  \begin{varwidth}{.65\textwidth}
    \centering
    \fbox{\subfloat{\begin{tikzpicture}[shorten >=1pt,->,font=\small]
  \tikzstyle{none}=[circle,minimum size=14pt,inner sep=0pt]
  \tikzstyle{final}=[circle,draw,style=double,minimum size=14pt,inner sep=0pt]

  \node[none]   (nn) at (0.2,0)     {};
  \node[final]  (l0) at (1,0)       {$l_0$};
  \node[final]  (l1) at (4,0)       {$l_1$};
  \node[final]  (l2) at (7,0)       {$l_2$};
  \node[above]  (l0r) at (l0.north) {$\scriptstyle\emptyset$};
  \node[above]  (l1r) at (l1.north) {$\scriptstyle\emptyset$};
  \node[above]  (l2r) at (l2.north) {$\scriptstyle\emptyset$};
  \node[none]   (H)  at (7.75,0.2)  {\textcolor{blue}{$\hypo_1$\ \ }};

  \draw (nn) -- (l0);
  
  \path (l0) edge[] node[above,yshift=-0.35em] {
    $\strans{\push(p)}{true}{-}$} (l1);
  
  \path (l1) edge[] node[above,yshift=-0.35em] {
    $\strans{\push(p)}{true}{-}$} (l2);

\end{tikzpicture}}}

    \fbox{\subfloat{\begin{tikzpicture}[shorten >=1pt,->,font=\small]
  \tikzstyle{none}=[circle,minimum size=14pt,inner sep=0pt]
  \tikzstyle{final}=[circle, draw,style=double,minimum size=14pt,inner sep=0pt]
  
  \node[none]  (nn) at (0.2,0)    {};
  \node[final] (l0) at (1,0)      {$l_0$};
  \node[final] (l1) at (4,0)      {$l_1$};
  \node[final] (l2) at (7,0)      {$l_2$};
  \node[none]  (H)  at (7.75,0.2) {\textcolor{blue}{$\hypo_2$\ \ }};
	
  \node[above] (l0r) at (l0.north) {$\scriptstyle\emptyset$};
  \node[above] (l1r) at (l1.north) {$\scriptstyle\{x_1\}$};
  \node[above] (l2r) at (l2.north) {$\scriptstyle\emptyset$};
  
  \draw (nn) -- (l0);
  
  \path (l0) edge[bend left=10] node[above] {
    $\strans{\push(p)}{true}{x_1 := p}$} (l1);
  \path (l1) edge[bend left=10] node[below] {
    $\strans{\pop(p)}{p = x_1}{-}$} (l0);
  
  \path (l1) edge[] node[above] {
    $\strans{\push(p)}{true}{-}$} (l2);
	
\end{tikzpicture}}}
  \end{varwidth}
  \caption{Three hypotheses constructed by \SLlambda: $\hypo_0$ (left), $\hypo_1$ and $\hypo_2$ (right).}
  \label{fig:demo:stack:01}
\end{figure}

This hypothesis is then subject to validation. Assume that it finds the counterexample $\push(0)\push(1)\push(2)$, which is accepted by $\hypo_0$ but rejected by
$\langstack$. Analysis of this counterexample reveals that $\epsilon$ and $\push(0)$ are inequivalent, since they are separated by the suffix $\push(p_1)\push(p_2)$
(since the concatenation of $\epsilon$ and $\push(p_1)\push(p_2)$ is accepted for all $p_1,p_2$ but $\push(0)\cdot \push(1)\push(2)$ is always rejected for all $p_1,p_2$).
It could seem natural to add $\push(p_1)\push(p_2)$ to~$\symbsuffs$, but $\SLlambda$ will not do that, since it follows the principle (from $L^\lambda$~\cite{Llambda})
that a new prefix in $S_p$ must extend an existing prefix by one symbol, and that a new suffix in $\symbsuffs$ must prepend one symbol to an existing one.
This principle keeps $S_p$ prefix-closed and $\symbsuffs$ suffix-closed, and aims to avoid inclusion of unnecessarily long sequences.
Therefore, instead of adding $\push(p_1)\push(p_2)$ as a suffix, $\SLlambda$ enters the prefix $\push(0)$ into $S_p$, and adds
one-symbol extensions of $\push(0)$,  in this case $\push(0)\push(1)$ and $\push(0)\pop(1)$, to $\prefixes$.
It notes that $\push(0)\push(1)$ is inequivalent to both $\epsilon$ and $\push(0)$, separated by the suffix $\push(p_1)$.
Again, $\push(0)\push(1)$ is therefore promoted to a short prefix, and its one-symbol extensions, $\push(0)\push(1)\push(2)$ and $\push(0)\push(1)\pop(2)$, are entered into~$\prefixes$.
Now, \SLlambda is able to add suffixes to $\symbsuffs$ that separate all prefixes in $S_p$, by two operations that achieve consistency.
\begin{enumerate}
\item
The $\push$-extensions of $\epsilon$ and $\push(0)\push(1)$, (i.e., $\push(0)$ and $\push(0)\push(1)\-\push(2)$)
are separated by the empty suffix, hence these
two prefixes are separated by the suffix $\push(p_1)$,
a one-symbol extension of $\epsilon$
which is added to $\symbsuffs$.
\item
The $\push$-extensions of $\epsilon$ and $\push(0)$  (i.e., $\push(0)$ and $\push(0)\push(1)$)
are separated by the suffix $\push(p_1)$, hence $\epsilon$ and $\push(0)$
are separated by $\push(p_1)\push(p_2)$, formed by prepending a symbol to the just added suffix $\push(p_1)$, which is added to $\symbsuffs$.
\end{enumerate}
After adding the suffixes, the closedness and consistency criteria are met, producing hypothesis $\hypo_1$ in \cref{fig:demo:stack:01}(right, top).
Assume that the validation of $\hypo_1$ finds counterexample $\push(0)\pop(0)$, which is in $\langstack$, but rejected by $\hypo_1$.
This counterexample reveals that after $\push(0)$, the two continuations $\pop(0)$ and $\pop(1)$ lead to inequivalent locations (separated by suffix $\epsilon$),
suggesting to refine the $\pop(p)$-transition after $\push(0)$. To this end, $\symbsuffs$ is extended by a suffix formed by prepending $\pop(p)$ to the empty suffix, and
a tree query is invoked for $\sdtfunction{\push(0)}{\pop(p_1)}$, which is $+$ iff $p_1 = x_1$ and $-$ otherwise. Since $\sdtfunction{\push(0)}{\pop(p_1)}$ makes a test for
$x_1$, which represents the data value of $\push(0)$, we infer that the data parameter of the $\push(0)$-prefix must be remembered in a register, and
that the $\pop(p)$-transition  must be split into two with guards $(x_1 \neq p)$ and $(x_1 = p)$.
The resulting hypothesis, $\hypo_2$, is shown in \cref{fig:demo:stack:01}(right, bottom), which is subject to another round of validation; the subsequent hypothesis construction reveals
the $\pop$-transitions from $l_2$ in \cref{fig:demo:stack}.

In \SLlambda, the sets $\prefixes$ and $\symbsuffs$ are maintained in a
\emph{classification tree} $CT$, a data structure that is specially designed to represent how the suffixes in $\symbsuffs$ partition $\prefixes$ into equivalence classes corresponding to locations.
This permits an optimization that can elide superfluous membership queries.
A classification tree is a decision tree. Each leaf is labeled by a subset of $\prefixes$.
Each inner node is labeled by a symbolic suffix $\symbsuff$ and induces a subtree for each equivalence class w.r.t.\ $\symbsuff$, whose leaves
contain prefixes in this equivalence class.
For example, in \cref{fig:demo:stack:ct}, which shows a $CT$ corresponding to hypothesis $\hypo_1$,
the nodes are labeled by the suffixes $\epsilon$, $\push(p_1)$ and $\push(p_1)\push(p_2)$,
which separate the leaves into four equivalence classes corresponding to the four locations in \cref{fig:demo:stack}.
Each edge is labeled by the results of the tree queries for a prefix in its equivalence class and the symbolic suffix of the source node.

\begin{figure}[t]
  \centering
  \begin{lrbox}{\epsr}
	\begin{tikzpicture}[edgegraph]
		\node[astate] {};
	\end{tikzpicture}
\end{lrbox}

\begin{lrbox}{\epsl}
	\begin{tikzpicture}[edgegraph]
		\node[state] {};
	\end{tikzpicture}
\end{lrbox}

\begin{lrbox}{\sbtrueacc}
	\begin{tikzpicture}[edgegraph]
		\node[ustate] at (0,0) (l0) {};
		\node[astate] at (1.6,0) (l1) {};
		
		\path (l0) edge[] node [above] {$true$} (l1);
	\end{tikzpicture}
\end{lrbox}

\begin{lrbox}{\sbtruerej}
	\begin{tikzpicture}[edgegraph]
		\node[ustate] at (0,0) (l0) {};
		\node[state] at (1.6,0) (l1) {};
		
		\path (l0) edge[] node [above] {$true$} (l1);
	\end{tikzpicture}
\end{lrbox}

\begin{lrbox}{\sbtruetrueacc}
	\begin{tikzpicture}[edgegraph]
		\node[ustate] at (0,0) (l0) {};
		\node[ustate] at (1.6,0) (l1) {};
		\node[astate] at (3.2,0) (l2) {};
		
		\path (l0) edge[] node [above] {$true$} (l1);
		\path (l1) edge[] node [above] {$true$} (l2);
	\end{tikzpicture}
\end{lrbox}

\begin{lrbox}{\sbtruetruerej}
	\begin{tikzpicture}[edgegraph]
		\node[ustate] at (0,0) (l0) {};
		\node[ustate] at (1.6,0) (l1) {};
		\node[state] at (3.2,0) (l2) {};
		
		\path (l0) edge[] node [above] {$true$} (l1);
		\path (l1) edge[] node [above] {$true$} (l2);
	\end{tikzpicture}
\end{lrbox}

\begin{tikzpicture}[shorten >=1pt,->,font=\scriptsize]
  \node[dtnode] (eps) at (0,5) {$\epsilon$};
  \node[dtnode] (push) at (1.5,4) {$\push(p_1)$};
  \node[dtnode] (pushpush) at (3.5,3) {$\push(p_1)\,\push(p_2)$};
  \node[align=left] (l0) at (-1.5,3) {
    $\underline{\pop(0)}$ \\
    $\pop(0)\push(1)$ \\
    $\pop(0)\pop(1)$ \\
    $\push(0)\pop(1)$ \\
    $\push(0)\push(1)\pop(2)$
  };
  \node[align=left,xshift=2em] (l1) at (4.5,2) {
    $\underline{\epsilon}$
  };
  \node[align=left] (l2) at (2.2,2) {
    $\underline{\push(0)}$
  };
  \node[align=left] (l3) at (0.65,3) {$\underline{\push(0)\push(1)}$};
  
  \path (eps)  edge[] node[right,yshift=.5em,align=center] {\usebox{\epsr}} (push);
  \path (eps)  edge[] node[left,yshift=1em,align=center]  {\usebox{\epsl}} (l0.north);
  \path (push) edge[] node[right,yshift=.5em,align=center] {\usebox{\sbtrueacc}} (pushpush.north);
  \path (pushpush) edge[] node[right,yshift=.3em,align=center]  {\usebox{\sbtruetrueacc}} (l1.north);
  \path (push) edge[] node[left,yshift=.5em,align=center] {\usebox{\sbtruerej}} (l3.north);
  \path (pushpush) edge[] node[left,xshift=-.5em,yshift=.3em,align=center] {\usebox{\sbtruetruerej}} (l2.north);
\end{tikzpicture}
  \caption{Classification tree for hypothesis $\hypo_1$ in \cref{fig:demo:stack:01}. Short prefixes are underlined.\bjcom{I suggest to simply put + or -  on edges}}
  \label{fig:demo:stack:ct}
\end{figure}

Each tree query requires a number of membership queries which may grow exponentially with the length of the suffix. $\SLlambda$ reduces this number by \emph{restricting} the
involved symbolic suffix to induce fewer membership queries, as long as the tree query can still make the separation between prefixes or transitions for which it was invoked.
To illustrate, recall that the analysis of the counterexample $\push(0)\push(1)\push(2)$ for $\hypo_0$ shows that $\epsilon$ and $\push(0)$ are inequivalent.
To separate these, we need not na\"ively use the symbolic suffix $\push(p_1)\push(p_2)$; but we can restrict it by considering only values of $p_1$ and $p_2$ that
are \emph{fresh}, i.e., different from all other preceding parameters in the prefix and suffix. With this restriction, the suffix can still separate $\epsilon$ and $\push(0)$,
and the tree query for prefix $\push(0)$ requires only one membership query instead of five.

\section{Data Languages and Register Automata}
\label{sec:dlra}

In this section, we review background concepts on data languages and register automata.
Our definitions are parameterized on a \emph{theory}, which is
a pair $\tuple{\domain, \binrelations}$
where $\domain$ is a (typically infinite) domain of {\em data values}, and
$\binrelations$ is a set of {\em relations} (of arbitrary arity) on $\domain$.
Examples of theories include:
\begin{inparaenum}[(i)]
\item$\tuple{\mathbb{N}, \{=\} }$, the theory of natural numbers with equality,
  and
\item
  $\tuple{\mathbb{R}, \{ < \} }$, the theory of real numbers with inequality;
  this theory also allows to express equality between elements.
\end{inparaenum}
Theories can be extended with constants (allowing, e.g., theories of sums with constants).

\beginparagraph{Data Languages.}
We assume a set $\Sigma$ of {\em actions}, each with an arity that determines how
many parameters it takes from the domain $\domain$.
For simplicity, we assume that all actions have arity $1$; our techniques can be extended to handle
actions with arbitrary arities.
A {\em data symbol} is a term of form $\act(\dval)$,
where $\act$ is an action and $\dval \in \domain$ is a data value.
A {\em data word} (or simply \emph{word}) is a finite sequence of data symbols.
The concatenation of two words $u$ and $v$ is denoted
$uv$, often we then refer to $u$ as a \emph{prefix} and $v$ as a \emph{suffix}.
Two words
$\dword = \act_1(\dval_1) \ldots \act_n(\dval_n)$
and
$\ddword = \act_1(\dval_1') \ldots \act_{n}(\dval_{n}')$
with the same sequences of actions
are \mbox{\emph{$\binrelations$-indistinguishable}},
denoted $\dword \approx_{\binrelations} \ddword$, if
$R(\dval_{i_1},\ldots,\dval_{i_j}) \Leftrightarrow R(\dval_{i_1}', \ldots, \dval_{i_j}')$
whenever $R$ is a $j$-ary relation in $\binrelations$ and
$i_1, \cdots , i_j$ are indices among $1 \ldots n$.
A {\em data language} $\Lang$ is a set
of data words that respects $\binrelations$ in the sense that
$\dword \approx_{\binrelations} \ddword$ implies
$\dword \in \Lang \Leftrightarrow \ddword \in \Lang$.
We often represent data languages as mappings from the set of
words to $\set{+,-}$, where $+$ stands for {\em accept} and $-$ for {\em reject}.

\beginparagraph{Register Automata.}
We assume a set of {\em registers} $x_1, x_2, \ldots$, and a set of \emph{formal parameters} $p, p_1, p_2, \ldots$.
A {\em parameterized symbol} is a term of form
$\act(p)$, where $\act$ is an action and $p$ a {\em formal parameter}.
A \emph{constraint} is a conjunction of negated and unnegated relations (from $\binrelations$) over
registers and parameters.
An \emph{assignment} is a parallel update of registers
with values from registers or the formal parameter~$p$.
We represent it as a mapping $\remap$ from $\set{x_{i_1}, \ldots ,x_{i_m}}$ to
$\set{x_{j_1}, \ldots ,x_{j_n}} \cup \set{p}$,
meaning that the value $\remap(x_{i_k})$ is assigned to $x_{i_k}$, for $k = 1, \ldots , m$.
In multiple-assignment notation, this would be written
\(
x_{i_1}, \ldots ,x_{i_m}  := \remap(x_{i_1}), \ldots ,\remap(x_{i_m})
\).

\begin{definition}%
\label{def:ra}
A {\em register automaton} (RA) is a tuple
$\auto = (\locs, \initloc, \vars, \transitions, \lambda)$, where
\begin{itemize}
\item $\locs$ is a finite set of {\em locations}, with $\initloc\in\locs$ as the {\em initial location},
\item $\vars$ maps each location $\loc \in  \locs$ to a finite set
$\varsof{\loc}$ of registers,
\item $\transitions$ is a finite set of {\em transitions}, each of form
  $\tuple{\loc,\act(p),g,\remap,\loc'}$, where
\begin{itemize}
\item $\loc \in \locs$ is a source location and
$\loc' \in \locs$ is a target location,
\item $\act(p)$ is a parameterized symbol, %
\item $g$, the \emph{guard}, is a constraint over $p$ and $\varsof{\loc}$, and
\item $\remap$ (the {\em assignment}) is a mapping from $\varsof{\loc'}$
   to $\varsof{\loc} \cup \set{p}$,
   and
\end{itemize}
\item $\lambda$ maps each $\loc \in \locs$ to $\set{+,-}$,
where $+$ denotes {\em accept} and $-$ {\em reject}.
\hfill\qed
\end{itemize}
\end{definition}

A \emph{state} of a RA
$\auto = (\locs, \initloc, \vars, \transitions, \lambda)$
is a pair $\tuple{\loc,\valuation}$
where $\loc \in \locs$ and $\valuation$ is a valuation over $\varsof{\loc}$,
i.e., a mapping from $\varsof{\loc}$ to $\domain$.
A \emph{step} of $\auto$, denoted
$\tuple{\loc,\valuation} \xrightarrow{\act(\dval)}
\tuple{\loc',\valuation'}$, transfers the state of $\auto$
from $\tuple{\loc,\valuation}$
to $\tuple{\loc',\valuation'}$ on input of the data symbol $\act(\dval)$
if there is a transition
$\tuple{\loc,\act(p),g,\remap,\loc'} \in \transitions$ such that
\begin{inparaenum}[(i)]
 \item $\valuation ~\models~ g[\dval/p]$, i.e., $d$ satisfies the guard $g$
under the valuation $\valuation$, and
 \item $\valuation'$ is defined by
$\valuation'(x_i) = \valuation(x_j)$ if $\remap(x_i)= x_j$, otherwise
$\valuation'(x_i) = d$ if $\remap(x_i) = p$.
\end{inparaenum}
A \emph{run} of $\auto$ over a data word
$\dword = \act(\dval_1)\ldots\act(\dval_n)$
is a sequence of steps of $\auto$
\[\tuple{\initloc,\valuation_0} ~ \xrightarrow{\act_1(\dval_1)} ~
\tuple{\loc_1,\valuation_1} \quad
\ldots \quad
\tuple{\loc_{n-1},\valuation_{n-1}} ~ \xrightarrow{\act_n(\dval_n)} ~
\tuple{\loc_{n},\valuation_n}
\]
for some initial valuation $\valuation_0$.
The run is {\em accepting} if $\lambda(\loc_n) = +$
and {\em rejecting} if $\lambda(\loc_n) = -$. The word $\dword$ is
{\em accepted (rejected) by $\auto$ under $\valuation_0$}
if $\auto$ has an accepting (rejecting) run over $\dword$
from $\tuple{\initloc,\valuation_0}$.
Define the language $\Langof{\auto}$ of $\auto$ as the set of words accepted by $\auto$.
A language is \emph{regular} if it is the language of some RA.

We require a RA to be \emph{determinate}, meaning that
there is no data word over which it has
both accepting and rejecting runs.
A determinate RA can be easily transformed into a deterministic one by strengthening its guards, and a deterministic
RA is by definition also determinate.
Our construction of RAs in \cref{sec:algorithm}
will generate determinate RAs which are not necessarily deterministic.
RAs have been extended to Register Mealy Machines (RMM) in several works
and it has been established how RA learning algorithms can be used
to infer models of systems with inputs and
outputs~\cite{Cassel2015RalibLearnlibExtension}, which we do, too.

\section{The $SL^\lambda$ Learning Algorithm}
\label{sec:algorithm}

In this section, we present the main building blocks of $\SLlambda$
before an overview of the main algorithm,
followed by techniques for reducing the cost of tree queries (page~\pageref{sec:CSS})
and for analyzing counterexamples (page~\pageref{sec:AnalysisCE}).

\beginparagraph{Symbolic Decision Trees.}
The functions, of form $\sdtfunction{u}{\symbsuff}$, that result from tree queries, should represent how the language $\Lang$ to be learned processes instantiations of $\symbsuff$ after the prefix $u$. Since $\SLlambda$ is intended to construct canonical RAs, it is natural to let these functions have the form of a tree-shaped ``mini-RA'', which we formalize as
\emph{symbolic decision trees} of a certain form.

For a word $u = \act_1(\dval_1)\ldots \act_k(\dval_k)$ and a symbolic suffix
$\symbsuff = \act_1'(p_1)\ldots \act_m'(p_m)$
a {\em $(u,\symbsuff)$-path} $\upath$ is a sequence $\vect gm$, where
each $g_i$ is a constraint over $\vect xk$ and $\vect pi$.
Define the condition represented by $\upath$, denoted $\pathcondof{\upath}$, as $g_{1} \land \cdots \land g_{m}$.
A {\em $(u,\symbsuff)$-tree} $\uvtree$ is a mapping from  a
set $\domof{\uvtree}$ of $(u,\symbsuff)$-paths to $\set{+,-}$.
Write $\dvaloverbar$ for $\vect{\dval}{k}$, $\dvalprimeoverbar$ for $\vect{\dval'}{m}$, $\xoverbar$ for $\vect xk$ and $\poverbar$ for $\vect pm$.
A $(u,\symbsuff)$-tree $\uvtree$ can be seen a function
with parameters $\xoverbar,\poverbar$ to $\set{+,-}$,
defined by $\uvtree(\xoverbar,\poverbar) = \uvtree(\upath)$
whenever $\upath \in \domof{\uvtree}$ and $\pathcondof{\upath}(\xoverbar,\poverbar)$ holds.
That is, for data values $\dvaloverbar$ and $\dvalprimeoverbar$ and each $(u,\symbsuff)$-path $\upath$, we have
$\uvtree(\dvaloverbar,\dvalprimeoverbar) = \uvtree(\upath)$ whenever $\pathcondof{\upath}(\dvaloverbar,\dvalprimeoverbar)$ is true.
If $\Lang$ is a data language, then $\sdtfunction{u}{\symbsuff}$ is a $(u,\symbsuff)$-tree representing membership in $\Lang$ in the sense that
for any values of $\vect pm$ we have $\sdtfunction{u}{\symbsuff}(\dvaloverbar,\poverbar) = +$ iff  $u\act_1'(p_1)\ldots \act_m'(p_m) \in \Lang$, and
$\sdtfunction{u}{\symbsuff}(\dvaloverbar,\poverbar) = -$ iff $u\act_1'(p_1)\ldots \act_m'(p_m) \not\in \Lang$.
For example,~\cref{fig:sdt:push-pop} shows a $(u, \symbsuff)$-tree where $u=\push(\dval_1)$ and $\symbsuff=\push(p_1)\pop(p_2)$.
This tree maps the $(u, \symbsuff)$-path $\true \land p_2=p_1$ to $+$ and $\true \land p_2\neq p_1$ to $-$.
From this, we can determine, \eg, that the word $\push(0)\push(1)\pop(1) \in \Lang$, but $\push(0)\push(1)\pop(2) \not\in \Lang$.

\SLlambda generates $(u,\symbsuff)$-trees $\sdtfunction{u}{\symbsuff}$
representing the language $\Lang$ to be learned through
so-called \emph{tree queries},
which perform membership queries for values of the data parameters $\vect pm$ that cover relevant equivalence classes of $\approx_{\binrelations}$.

From the results of tree queries, we can extract registers and guards in the location reached by a prefix $u$. Intuitively, the registers must remember the data values of $u$ that occur in some guard in some $\sdtfunction{u}{\symbsuff}$, and the outgoing guards from the location reached by
$u$  can be derived from the initial guards in the trees $\sdtfunction{u}{\symbsuff}$, since the initial guards represent the constraints that are used when processing the first symbol of $\symbsuff$.
Let $\getmemorable{u}{\treeoracle}{\symbsuff}$, the set of {\em memorable parameters}, denote the set of
registers among $\set{\vect xk}$ that occur on some $(u,\symbsuff)$-path in
$\domof{\sdtfunction{u}{\symbsuff}}$. Intuitively, if $x_i$ is a memorable parameter, then the $i^{th}$ data value in $u$ will
be remembered in the register $x_i$ in the location reached by $u$.
For example, for the stack in~\cref{fig:demo:stack}, in the location reached by $\push(0)$ the data value $\dval_1=0$ is memorable so will be remembered in register $x_1$.
Note that a $(u,\symbsuff)$-tree itself does not have any registers: it only serves to show which registers are needed in the location reached by $u$ in the
to-be-constructed automaton.
Define $\getmemorable{u}{\treeoracle}{\symbsuffs}$
as $\cup_{\symbsuff \in \symbsuffs} \getmemorable{u}{\treeoracle}{\symbsuff}$.
For a prefix $u$ and symbolic suffix $\csuff$ whose first action is $\act$, let $\guardsof{u}{\set{\csuff}}{\alpha}$ denote the
initial guards in the $(u,\symbsuff)$-tree $\sdtfunction{u}{\symbsuff}$, with
$p_1$ replaced by $p$.
For a set~$\symbsuffs$, let $\guardsof{u}{\symbsuffs}{\alpha}$ denote the set of satisfiable conjunctions of guards in
$\guardsof{u}{\set{\csuff}}{\alpha}$ for $\csuff \in \symbsuffs$ with first action~$\act$.

Two $(u,\symbsuff)$-trees, $\uvtree$ and $\uvtree'$, are \emph{equivalent}
denoted $\uvtree \equiv \uvtree'$, if
$\domof{\uvtree} = \domof{\uvtree'}$ and
$\uvtree(\upath) = \uvtree'(\upath)$ for each $\upath \in \domof{\uvtree}$.
For a mapping $\gamma$ on registers, we define its extension to
$(u,\symbsuff)$-paths in the natural way. For a $(u,\symbsuff)$-tree $\uvtree$, we define
$\gamma(\uvtree)$ by
$\domof{\gamma(\uvtree)} = \setcomp{\gamma(\upath)}{\upath \in \domof{\uvtree}}$
and $\gamma(\uvtree)(\gamma(\upath)) = \uvtree(\upath)$.

Let $u \plainequiv{\treeoracle}{\symbsuffs} u'$ denote that
$\sdtfunction{u}{\symbsuff}\equiv \sdtfunction{u'}{\symbsuff}$,
for all symbolic suffixes $\symbsuff \in \symbsuffs$.
Let $u \remapequiv{\treeoracle}{\symbsuffs}{\gamma} u'$ denote that
$\gamma$ is a bijection from
$\getmemorable{u}{\treeoracle}{\symbsuffs}$ to $\getmemorable{u'}{\treeoracle}{\symbsuffs}$ such that for all $\symbsuff \in \symbsuffs$ we have
$\gamma(\sdtfunction{u}{\symbsuff})\equiv \sdtfunction{u'}{\symbsuff}$.
Let $u \remapequiv{\treeoracle}{\symbsuffs}{} u'$ denote that
$u \remapequiv{\treeoracle}{\symbsuffs}{\gamma} u'$ for some bijection $\gamma$.
Intuitively, two words $u$ and $u'$ are equivalent if there is a bijection $\gamma$
which for each $\symbsuff \in \symbsuffs$
transforms $\sdtfunction{u}{\symbsuff}$ to $\sdtfunction{u'}{\symbsuff}$,
Note that in general, when $u \remapequiv{\treeoracle}{\symbsuffs}{} u'$,
there can be several such bijections.

\beginparagraph{Data Structures.}
During the construction of a hypothesis, the \SLlambda algorithm maintains:
\begin{inparaenum}[(i)]
\item a prefix-closed set $S_p$ of \emph{short prefixes}, representing locations,
\item and a set of one-symbol extensions of the prefixes in $S_p$, representing transitions;
  we use $\prefixes$ to represent the union of $S_p$ and this set, and
\item a suffix-closed set $\symbsuffs$ of symbolic suffixes.
\end{inparaenum}
Each one-symbol extension of form $u\dact{\act}{\dval}$ is formed to let $\dval$ satisfy a specific guard $g$; we then always
choose $\dval$ as a {\em representative data value}, denoted $\repr ug$, satisfying $g$ after $u$.

The sets $\prefixes$ and $\symbsuffs$ are maintained in a
\emph{classification tree} $CT$, which is designed to represent
how the suffixes in $\symbsuffs$ partition the set $\prefixes$ into equivalence classes corresponding to locations.
A classification tree is a rooted tree, consisting of nodes connected by edges. Each inner node
is labeled by a symbolic suffix, and each leaf is labeled by a subset of $\prefixes$.
To each node $\ctnode$ is assigned a representative prefix $\rapof{\ctnode}$ in $\prefixes$.
For a node $\ctnode$, let $\symbsuffof{\ctnode}$ its suffix and $\ancestorsof{\ctnode}$ denote the set of symbolic suffixes of $\ctnode$ and all its ancestors in the tree.
Each outgoing edge from~$\ctnode$ corresponds to an equivalence class of $\remapequiv{\treeoracle}{\ancestorsof{\ctnode}}{}$ from which a representative member
is chosen as the representative prefix of its target node.
Each leaf node~$\leafnode$ is labeled by a set of data words, which are all in the same equivalence class of $\remapequiv{\treeoracle}{\ancestorsof{\leafnode}}{}$.
Thus, nodes in different leaves are guaranteed to be inequivalent, since they are separated by the symbolic suffixes in $\ancestorsof{\lcaof{\leafnode}{\leafnode'}}$, where
$\lcaof{\leafnode}{\leafnode'}$ is the lowest common ancestor node of $\leafnode$ and $\leafnode'$.
We let $\prefixmap$ denote the mapping, which maps each prefix $u \in \prefixes$ to the classification tree leaf where it is contained.
We also let $\ancestorsof{u}$ denote $\ancestorsof{\prefixmap(u)}$, the suffixes of all ancestors of $\prefixmap(u)$.
The representative prefix, $\rapof{\leafnode}$, of each leaf node $\leafnode$ will induce a location in the RA to be constructed.

\begin{algorithm}[t!]
  \DontPrintSemicolon
  \SetKwComment{step}{\(\triangleright\)  }{}
  \SetKwBlock{Let}{let}{end}
  \SetKwBlock{Loop}{loop}{end loop}
  \SetKwProg{Fn}{Function}{ is}{end}
  \caption{Operations on the Classification Tree.}
  \label{alg:sub-routines}

  \Fn{$\Siftfn(u, \ctnode)$}{
    \lIf{$\ctnode$ \mbox{\rm is a leaf}}{$\prefixmap \gets \prefixmap[u \mapsto \ctnode]$
    } \uElse {
      Compute $\sdtfunction{u}{\symbsuffof{\ctnode}}$\;
      \lIf{$\ctnode$ has child $\ctnode'$ 
  	with $u \remapequiv{\treeoracle}{\ancestorsof{\ctnode}}{} \rapof{\ctnode'}$}{$\Siftfn(u, \ctnode')$
      } \uElse {
        Create new leaf $\ctnode'$ as child of $\ctnode$ with $\rapof{\ctnode'} = u$\;
        $\Siftfn(u, \ctnode')$\;
      }
    }	
  }
  \;\vspace*{-.35em}
  \Fn{Expand($u$)}{
    $S_p \leftarrow S_p \cup \{u\}$\;
    \lFor{$\alpha \in \acts$}{%
      $\Siftfn(u\dact{\act}{\repr ug}, \rootof{CT})$ for each $g\in\guardsof{u}{\ancestorsof{u}}{\alpha}$
    }  
  }
  \;\vspace*{-.35em}
  \Fn{Refine($\ctnode$, $\symbsuff$)}{
    Replace $\ctnode$ by an inner node $\ctnode'$ with $\symbsuffof{\ctnode'} = \symbsuff$\;
    \lFor{$u \in \invU(\ctnode)$}{$\Siftfn(u, \ctnode')$}
  }

\end{algorithm}

The insertion of a new prefix $u$ into the classification tree $CT$ is performed by function $\Siftfn$ (cf.~\cref{alg:sub-routines}).
It traverses the $CT$ from the root downwards.
At each internal node $\ctnode$, it checks whether $u \remapequiv{\treeoracle}{\ancestorsof{\ctnode}}{} \rapof{\ctnode'}$ for any child $\ctnode'$ of~$\ctnode$. If so, it continues the traversal at $\ctnode'$, otherwise a
new child of $\ctnode$ is created as a leaf $\leafnode$ with $\rapof{\leafnode} = u$.
When reaching a leaf $\leafnode$, the mapping $\prefixmap$ is updated to reflect that $u$ has been sifted to $\leafnode$.
In the classification tree in Fig.~\ref{fig:demo:stack:ct}, e.g.,
$\epsilon$ is the representative prefix
of inner nodes $\push(p_1)$ and $\push(p_2)\push(p_2)$
as it is the first prefix that was sifted down this path.
The short prefix $\push(0)$ at the second leaf from right
was sifted from the root to $\push(p_1)$
and then to $\push(p_1)\push(p_2)$ as
$\push(0) \remapequiv{\treeoracle}{ \set{\epsilon} }{} \epsilon$
and
$\push(0) \remapequiv{\treeoracle}{ \set{\epsilon,~\push(p_1)} }{} \epsilon$.
Since, however,
$\push(0) \not\remapequiv{\treeoracle}{ \mathcal{V} }{} \epsilon$
for $\mathcal{V}= \set{\epsilon,~\push(p_1), \push(p_1)\push(p_2)}$, a new leaf was created and $\push(0)$ was made the representative
prefix of the new leaf and a short prefix.

\begin{algorithm}[t!]
  \linespread{1.2}\selectfont
  \DontPrintSemicolon
  \SetKwComment{step}{\(\triangleright\)  }{}
  \SetKwBlock{Loop}{loop}{end loop}
  \caption{\SLlambda Learning.}
  \label{alg:main}

{
  Initialize $CT$ as inner node $\rootof{CT}$ with suffix $\epsilon$ and $\prefixes \gets \emptyset$, $S_p \gets \emptyset$\;
 	$\Siftfn(\epsilon, \rootof{CT})$\;
 		\texttt{HYP:} \Repeat{closed and consistent}{
	    	\step{Check closedness}
	      \If(\tcp*[f]{location}){exists leaf $\leafnode$ for which $\invU(\leafnode) \cap S_p = \emptyset$}{
	        $\Expandfn(u)$ for some $u\in \invU(\leafnode)$\;
	      }
				\If(\tcp*[f]{transition}){$u\in S_p$ and $g\in\guardsof{u}{\ancestorsof{u}}{\alpha}$  but $u\dact{\act}{\repr ug} \notin \prefixes$}{
	        $\Siftfn(u\dact{\act}{\repr ug}, \rootof{CT})$\;
				}
        \If(\tcp*[f]{register}){$u\dact{\act}{\dval} \in \prefixes$ s.t.\ $\getmemorable{u\dact{\act}{\dval}}{\treeoracle}{\mathcal{V}\!(\!u\act\!(\!\dval\!)\!)}\! \not\subseteq \!
  \getmemorable{u}{\treeoracle}{\ancestorsof{u}}\!\cup\!\set{x_{\sizeof{u}\!+\!1}}\!$}{
          Let $\symbsuff \in \ancestorsof{ u\dact{\act}{\dval} }$ with $\getmemorable{u\dact{\act}{\dval})}{\treeoracle}{\symbsuff} \not\subseteq (\getmemorable{u}{\treeoracle}{\ancestorsof{u}} \cup \set{x_{\sizeof{u}+1}})$\;
          $\Refinefn(\prefixmap(u), \symbalpha \symbsuff)$\;
        }
				\step{Check consistency}        
				\If(\tcp*[f]{location}){$u, u' \in \invU(L) \cap S_p$ with $u \remapequiv{\treeoracle}{\ancestorsof{\leafnode}}{\gamma} u'$  for leaf $\leafnode$ 
	      with $u\dact{\act}{\repr ug}, u'\dact{\act}{\repr{u'}{\gamma(g)}}\in \prefixes$ 
				but $\prefixmap(u\dact{\act}{\repr ug}) \neq \prefixmap(u'\dact{\act}{\repr{u'}{\gamma(g)}})$
				} {
					$\Refinefn(\prefixmap(u), \symbalpha \symbsuff)$ with
					$\symbsuff = \symbsuffof{\lcaof{ u\dact{\act}{\repr ug}}{ u'\dact{\act}{\repr{u'}{\gamma(g)}}}}$\;
				}
				\If(\tcp*[f]{transition(a)}){$g\in\guardsof{u}{\ancestorsof{u}}{\alpha}$ and $u\dact{\act}{\repr ug}, u\dact{\act}{\dval} \in \prefixes$ with $(u,\dval) \vDash g$ but $\prefixmap(u\dact{\act}{\repr ug}) \neq \prefixmap(u\dact{\act}{\dval})$}{
					$\Refinefn(\prefixmap(u), \symbalpha \symbsuff)$ with       
					$\symbsuff = \symbsuffof{\lcaof{ u\dact{\act}{\repr ug} }{ u\dact{\act}{\dval} }}$\;
				}
				\If(\tcp*[f]{transition(b)}){$u\dact{\act}{\repr ug}, u\dact{\act}{\dval} \in \prefixes$ with $u\dact{\act}{\repr ug} \not\remapequiv{\treeoracle}{\ancestorsof{u\dact{\act}{\dval}}}{\mathbf{id}}  u\dact{\act}{\dval}$}{
					$\Refinefn(\prefixmap(u), \symbalpha \symbsuff)$ with    
					$\symbsuff$ s.t. $u\dact{\act}{\repr ug} \not\remapequiv{\treeoracle}{\set{\symbsuff}}{\mathbf{id}}  u\dact{\act}{\dval}$\;
				}	          
        \If(\tcp*[f]{register}){$u,u\act \in \prefixes$ with %
      $u \remapequiv{\treeoracle}{\ancestorsof{u}}{\gamma} u$
        and no extension $\gamma'$ of $\gamma$ with $u\dact{\act}{\dval} \remapequiv{\treeoracle}{\ancestorsof{u\dact{\act}{\dval}}}{\gamma'} u\dact{\act}{\dval}$}{
          $\Refinefn(\prefixmap(u), \symbalpha \symbsuff)$ with $\symbsuff$ s.t. $u\dact{\act}{\dval} \not\remapequiv{\treeoracle}{\set{\symbsuff}}{\gamma'} u\dact{\act}{\dval}$ for any $\gamma'$\;
        }
			}
			$\hypo \leftarrow$ Hypothesis($CT$)\;
			\lIf{$\exists w \in \Sigma^{+}\ s.t.\ \hypo(w) \neq \Lang(w)$}{$\Analyzefn(w)$ and \textbf{goto} \texttt{HYP} \textbf{else} \Return $\hypo$}
}

\end{algorithm}

\beginparagraph{The \SLlambda Algorithm.}
The core of the \SLlambda algorithm, shown in \cref{alg:main},
initializes the classification tree to consist
of one (root) inner node, for the empty suffix (which classifies words as accepted or
rejected); $\prefixes$ and $S_p$ are empty. It then sifts the empty prefix $\epsilon$, thereby entering it into $\prefixes$.
Thereafter, \cref{alg:main} repeats a main loop in which $CT$ is checked for a number of
closedness and consistency properties. Whenever such a property is not satisfied, a corrective update is made by adding information to $CT$.
These corrective updates fall into two categories, carried out by the following functions:
\begin{itemize}
\item
  $\Expandfn$ takes a prefix $u \in \prefixes$ and makes it into a short prefix. Since each short prefix must have a set of one-symbol
  extensions in $\prefixes$, the function forms one-symbol extensions
  of form $u\act(\repr ug)$, which are entered into the classification tree by sifting.
\item
$\Refinefn$ takes a leaf node $\leafnode$ and a symbolic suffix $\csuff$; it sifts the prefixes $u$ in $\leafnode$, thereby obtaining
$\sdtfunction{u}{\csuff}$ from a tree query. This can either split $\leafnode$ into several equivalence classes, refine
the initial guards or extend the set of registers in the location represented by $\leafnode$.
\end{itemize}
Let us now describe the respective corrective updates in \cref{alg:main}.

\littleparagraph{Location Closedness}
is satisfied if each leaf contains a short prefix in $S_p$. Whenever a leaf $\leafnode$ does not contain a short prefix in $S_p$, one of its
prefixes $u$ is chosen for inclusion in $S_p$ by calling $\Expandfn(u)$, which adds one-symbol extensions to $\prefixes$.

\littleparagraph{Transition Closedness}
is satisfied if for each short prefix $u$, action $\alpha$, and initial guard in
$\guardsof{u}{\ancestorsof{u}}{\alpha}$, the extension $u\dact{\act}{\repr ug}$ is in $\prefixes$. If this is not
satisfied, the missing $u\dact{\act}{\repr ug}$ is added to $\prefixes$ by sifting into $CT$.

\littleparagraph{Register Closedness}
is satisfied if for each pair of prefixes $u$ and $u\act(\dval)$ in $\prefixes$, the memorable parameters found for $u$ contain the memorable parameters revealed by the suffixes for $u\act(\dval)$, except for $x_{\sizeof{u}+1}$, where $\sizeof{u}$ is the length of $u$.
Register closedness guarantees that in a hypothesis $\hypo$, values of registers in the location of $u\act(\dval)$ can all be obtained by assignment from the registers in location $u$ and the
just received parameter.
If it is not satisfied, a suffix $\symbsuff$ for $u\act(\dval)$ which reveals a missing register is prepended by $\act(p_1)$ and added to the suffixes for $u$, whereafter $\Refinefn(\prefixmap(u), \symbalpha \symbsuff)$ will reveal the missing parameter. Here, and in the following, we use $\symbalpha$ to denote $\act(p_1)$, and $\symbalpha \symbsuff$ to denote the result $\act(p_1) \act_2'(p_2)\ldots \act_{m+1}'(p_{m+1})$ of prepending $\symbalpha$ to $\symbsuff = \act_1'(p_1)\ldots \act_m'(p_m)$.
If possible, we try to choose a shortest~$\symbsuff$, and also
restrict the parameters of $\symbalpha\symbsuff$ to reduce the cost of
the tree query for $\sdtfunction{u}{\symbalpha\symbsuff}$.

\littleparagraph{Location Consistency.}
Analogously to consistency in the classic $L^*$ algorithm,
we split a leaf containing two short prefixes $u$, $u'$, in case their
corresponding extensions are not equivalent, i.e.,
there is a $g \in \guardsof{u}{\ancestorsof{u}}{\alpha}$ such that $\prefixmap(u\act(\repr ug)) \neq \prefixmap(u'\act(\repr{u'}{\gamma(g)}))$.
The splitting is done by calling $\Refinefn(\prefixmap(u), \symbalpha \symbsuff)$, where $\symbsuff$ is the symbolic suffix labeling the
common ancestor of the leaves of $u\act(\repr ug)$ and $u'\act(\repr{u'}{\gamma(g)})$.

\littleparagraph{Transition Consistency}
is satisfied if %
all one-symbol extensions $u\alpha(\dval)$ that satisfy some guard $g$ in $\guardsof{u}{\ancestorsof{u}}{\alpha}$,
are sifted to the same leaf as the extension $u\alpha(\repr ug)$ with the representative data value $\repr ug$.
If not, the guard $g$ should be split by
calling $\Refinefn(\prefixmap(u), \symbalpha \symbsuff)$, where
$\symbsuff$ is the symbolic suffix labeling the
common ancestor of the leaves of $u\act(\repr ug)$ and $u\act(\dval)$.
A similar case (\textit{Transition Consistency(b)}) occurs when $u\alpha(\repr ug)$ and $u\alpha(\dval)$ are sifted to the same leaf, but are not equivalent under the
identity mapping between registers. Also here, the guard $g$ should be split by
calling $\Refinefn(\prefixmap(u), \symbalpha \symbsuff)$, where
$\symbsuff$ is a shortest suffix under which $u\act(\repr ug) \not\remapequiv{\treeoracle}{\set{\symbsuff}}{\mathbf{id}}  u\act(\dval)$.

\littleparagraph{Register Consistency.}
For some short prefix $u$ with
memorable values $\getmemorable{u}{\treeoracle}{\ancestorsof{u}}$,
there may be symmetries in $\sdtfunction{u}{\symbsuff}$ for some $\symbsuff \in \ancestorsof{u}$,
i.e., for some permutation~$\gamma$ on $\getmemorable{u}{\treeoracle}{\ancestorsof{u}}$
we have $u \remapequiv{\treeoracle}{\ancestorsof{u}}{\gamma} u$.
It may be that this symmetry does not exist in the SUL, but we did not yet add a suffix that disproves it \iffextendedver{(cf. example in~\cref{sec:symmetry})}.
Register consistency checks for the existence of such
suffixes by comparing symmetries in $u$ and its continuations $u\dact{\act}{\dval}$.
If a symmetry between data values of $u$
does not exist in $u\alpha$ while one or more of the data
values are memorable in $u\alpha$, we can construct a
suffix that breaks the symmetry also for $u$.

\beginparagraph{Restricted Symbolic Suffixes.}\label{sec:CSS}
To reduce the number of membership queries for tree queries of form $\sdtfunction{u}{\symbsuff}$, we impose, when possible,
restrictions to the parameters of $\symbsuff$, meaning that $\sdtfunction{u}{\symbsuff}$
 represents acceptance/rejection of $u\symbsuff$ \emph{only for the suffix parameters that satisfy the imposed restrictions}.
An illustration was given at the end of \cref{sec:ideas}.
A more detailed description
\iffwithappendix{is in \cref{app:suffix-optimization}}{appears
  in the extended version~\cite{SLlambda@arXiv-24} of this paper}.
Since a restricted symbolic suffix $\symbsuff'$ represents fewer actual suffixes than an unrestricted one $\symbsuff$, it has less separating power,
so suffixes should only be restricted if their separating power is sufficient.
The principles for adding restrictions are specific to the theory; we have implemented them for the theory $\tuple{\mathbb{N}, \{=\}}$.
There, we consider two forms of restrictions on suffix parameters $p_i$:
\begin{inparaenum}[(i)]
\item $\freshof{p_i}$, meaning that $p_i$ is different from all other preceding parameters in the prefix and suffix,
\item $p_i = p_j$, where $j< i$, i.e., $p_j$ is an earlier parameter in the restricted suffix.
\end{inparaenum}
Let us consider how restricted suffixes arise when prepending an action $\symbalpha$ to an existing suffix~$\symbsuff$, in a call of form $\Refinefn(\prefixmap(u), \symbalpha\symbsuff)$,
in the case that $u$, $\symbalpha$, and $\symbsuff$ are chosen such that
$\getmemorable{u\dact{\act}{\dval})}{\treeoracle}{\symbsuff}$ contains a particular memorable parameter.
Let us denote the parameters of $\symbalpha\symbsuff$ by $p_1,\cdots,p_{|\symbsuff|+1}$.
The restriction of suffix $\symbalpha \symbsuff$ is then obtained~by
\begin{enumerate}
\item
letting the parameter of $\symbalpha$ be fresh if $\dval$ is not equal to a previous data value in $u$, and
\item
restricting each parameter $p_i$ with $i >1$ in $\symbalpha\symbsuff$ to be
\begin{inparaenum}[(i)]
	\item fresh whenever $p_{i-1}$ is fresh in $\symbsuff$ or the branch taken in $\sdtfunction{u\dact{\act}{\dval}}{\symbsuff}$ for fresh $p_{i-1}$  reveals the sought register, and
	\item equal to a previous value $p_j$ in $\symbalpha\symbsuff$ if the branch taken in $\sdtfunction{u\dact{\act}{\dval}}{\symbsuff}$ for $p_{i-1}$ equal to the corresponding value reveals the sought register.
\end{inparaenum}
\end{enumerate}

\beginparagraph{Hypothesis Construction.}
We can construct a hypothesis from a closed
and consistent classification tree.
Location closedness ensures that every transition
has a defined source and target location,
transition closedness ensures that every
transition that is observed by the tree queries
we have performed so far, is represented by
a prefix, and register closedness ensures that
registers exist for all memorable data values
in corresponding locations.
Location consistency, transition consistency, and
register consistency ensure that we can construct
a unique (up to naming of locations and registers)
determinate register automaton eventhough there
may exist multiple short prefixes for one
location and symmetries betweeen memorable data values.

We construct the register automaton $\auto = (\locs, \initloc, \vars, \transitions, \lambda)$, where
\begin{itemize}
  \item $\locs$ is the set of leaves of $CT$, and $\initloc$ is the leaf containing the empty prefix $\emptyword$,
  \item $\vars$ maps each location $\loc \in  \locs$ to
  $\getmemorable{u}{\treeoracle}{CT}$, where $u$ is
  the representative short prefix of the leaf corresponding to $\loc$, and
  \item $\lambda(\loc)=+$ if the  leaf
  $\loc$ is in the accepting subtree of the root, else $\lambda(\loc)=-$.
  \item for every location $\loc$ with short prefix $u$, action $\act$, and guard $g$ in $\guardsof{u}{\ancestorsof{u}}{\alpha}$,
  there is a transition $\tuple{\loc,\act(p),g,\remap,\loc'}$, where
  \begin{itemize}
\item $\loc' = \prefixmap(u\dact{\act}{\repr ug})$ is the target location, and
\item $\remap$ (the {\em assignment}) is defined by $\gamma$
for which $u\dact{\act}{\repr ug} \remapequiv{\treeoracle}{\ancestorsof{u\dact{\act}{\repr ug}}}{\gamma} \rapof{u\dact{\act}{\repr ug}}$
  \end{itemize}
\end{itemize}

\begin{algorithm}[tb]
  \DontPrintSemicolon
  \SetKwComment{step}{\(\triangleright\)  }{}
  \SetKwBlock{Let}{let}{end}
  \SetKwBlock{Loop}{loop}{end loop}
  \SetKwProg{Fn}{Function}{ is}{end}
  \caption{Analyze Counterexample.}
  \label{alg:analyze-ce}

	\Fn{Analyze($w$)}{
	  \For{$\lengthof{w} \geq i > 0$}{
	    \For{$u\in As(w_{1:i-1})$}{
	      Let $u\dact{\act}{\repr ug} \in \prefixes$ represent the last transition of $w_{1:i}$ in $\hypo$\;
	      Let $\symbsuff = \actsof{w_{i+1:\lengthof{w}}}$ (or $\epsilon$ for $i=\lengthof{w}$)\;
	      \If(\tcp*[f]{location}){$u\dact{\act}{\repr ug} \not\remapequiv{\treeoracle}{\set{\symbsuff}}{} {u'}$ for all $u'\in As(w_{1:i})$}{
	        $\Expandfn(u\dact{\act}{\repr ug})$ and stop analysis of $w$\;
	      }
	      \If(\tcp*[f]{transition}){initial guard $g$ in $\treequerysul{u}{\symbalpha\symbsuff}$ but no $u\dact{\act}{\repr ug} \in \prefixes$}{
	        $\Siftfn(u\dact{\act}{\repr ug}, \rootof{CT})$ and stop analysis of $w$\;
	      }
	    }
	  }
	}

\end{algorithm}

\beginparagraph{Analysis of Counterexamples.}\label{sec:AnalysisCE}
When an equivalence query returns a counterexample $w$,
we process the counterexample as is shown in \cref{alg:analyze-ce}.
From right to left,
we split the counterexample at every index
into a location prefix $w_{1:i-1}$,
a transition prefix $w_{1:i}$, and a suffix $w_{i+1:|w|}$.
We use the location and transition prefixes
to find corresponding short prefixes $u$ and
prefixes $u\dact{\act}{\repr ug}$ by tracing
$w_{1:i-1}$ and $w_{1:i}$ on the hypothesis.
We write $As(w_{1:i})$ for the short prefix corresponding
to the location reached by $w_{1:i}$ in a hypothesis and
$\valsof{w}$ for the sequence of actions of $w$.
We can then distinguish two cases:
  (1) The word $u\dact{\act}{\repr ug}$ is inequivalent to all
  corresponding short prefixes for the suffix of the counterexample.
  In this case, we make $u\dact{\act}{\repr ug}$ a short prefix.
  (2) The tree query $\treequerysul{u}{\symbalpha\symbsuff}$
  shows a new initial guard. In this case, we add the corresponding (new)
  prefix $u\dact{\act}{\repr ug}$ to the set of prefixes.
If neither case applies, we continue with the next index.
Since $w$ is a counterexample, one of the cases will
apply for some index
(cf.~\cref{lemma:ce}).

\section{Correctness and Complexity} \label{sec:correctness}
Let us now briefly discuss the correctness and query complexity
of \SLlambda. The correctness arguments are analogous
to the arguments presented for other active
learning algorithms. One notable difference to \SLstar is
that \SLlambda establishes register consistency instead
of relying on counterexamples for distinguishing symmetric
registers. Proofs \iffextendedver{and a more detailed discussion}
can be found in
\iffwithappendix{\cref{app:correctness}}{the
  extended version~\cite{SLlambda@arXiv-24} of this paper}.

\begin{lemma}\label{lemma:ce}
A counterexample leads to a new short prefix or
to a new prefix.
\end{lemma}
This is a direct consequence of \cref{alg:analyze-ce}.
Using  a standard construction that leverages properties of
counterexamples (cf.~\cite{RiSh:inference,CHJS:faoc-16}),
it can be shown that one of the two cases in the algorithm
will trigger for some index of the counterexample.
As long as expanding (or sifting) new prefixes does not
trigger a refinement, the current counterexample can be analyzed
again, until a refinement occurs.

\Cref{lemma:ce} establishes progress towards a finite RA for a language $\Lang$.
Let $m$ be the length of the longest counterexample, $t$ the number of
transitions, $r$ the maximal number of registers at any location,
and $n$ the number of locations in the final model.
($t$ dominates both $n$ and $r$.)
\begin{theorem}
\label{thm:complexity}
\SLlambda infers a RA for regular data language $\Lang$ with
$O(t)$ equivalence queries and
$O(t^2 \, n^r + tmn \, m^m)$ membership queries for
sifting words and analyzing counterexamples.
\end{theorem}

$O(t)$ is an improvement over the worst case estimate
of $O(tr)$ equivalence queries for \SLstar~\cite{CHJS:faoc-16}.
\SLlambda also improves the worst case estimate for membership queries
for sifting to $O(t^2 \, n^r)$ from $O(t^2r \, n^r$)
for filling the table in~\SLstar.
For analyzing counterexamples, \SLlambda replaces $O(trm \, m^m)$
with $O(tmn \, m^m )$.

\section{Evaluation}
\label{sec:evaluation}

As mentioned, we have implemented the \SLlambda algorithm in the
publicly available RALib tool for learning register automata.
RALib already implemented the \SLstar algorithm~\cite{CHJS:faoc-16}
that uses an observation table as its data structure.
In order to evaluate the effect of analyzing counterexamples
as described in \cref{sec:algorithm}, we have also implemented the
\SLCT classification tree learning algorithm that uses the same
counterexample analysis technique as the \SLstar algorithm,
i.e., adding suffixes from counterexamples to the classification
tree directly.
We compare the performance of the \SLlambda algorithm against that of
\SLstar and \SLCT.
All models, the experimental setup, and infrastructure for executing
the experiments are available on the paper's artefact at
Zenodo~\cite{SLlambda-artifact@Zenodo} and updated versions on
GitHub\footnote{\url{https://github.com/LearnLib/ralib-benchmarking}}.

\medskip
\noindent
\textbf{Experimental Setup.}
We use two series of experiments: (1)~A black-box learning
setup with random walks for finding counterexamples on small models
from the Automata Wiki~\cite{DBLP:conf/birthday/NeiderSVK97} to
establish a baseline comparison with other results and to
evaluate the impact of using non-minimal counterexamples.
In these experiments, we verify with a model checker
that the inferred model is equivalent to the SUL
and we stop as soon as the correct model
is produced by a learning algorithm.
(2)~A white-box setup with a model checker for finding
short counterexamples to analyze the scalability of algorithms on
(2a)~$24$ consecutive hypotheses of the
\href{https://github.com/Mbed-TLS/mbedtls/releases/tag/v2.26.0}{Mbed TLS 2.26.0}
server,\footnote{%
We obtained these hypotheses by extending the machinery of
\href{https://github.com/assist-project/dtls-fuzzer}{DTLS-Fuzzer}~\cite{DTLS-Fuzzer@ICST-22},
a publicly available tool for learning state machine models of DTLS
implementations.}
as well as
(2b)~sets of randomly generated automata.\footnote{%
We used the algorithm of Champarnaud and
Paranthoën~\cite{Champarnaud2005RandomGenerationDfas}
to enumerate semantically distinct DFAs with a specific alphabet
and number of locations. We then replaced the alphabet symbols
with RA actions of arity one, and finally replaced a fraction of the
transitions with simple gadgets that store and compare
data values.}
All results were obtained on a MacBook Pro with an Apple M1 Pro CPU
and $32$~GB of memory, running macOS version $12.5.1$ and OpenJDK version
$17.0.8.1$.

\begin{table}[t!]
\caption{Results on AutomataWiki Systems.}
\label{tbl:wikiresults}
\centering
\resizebox{.99\textwidth}{!}{
\begin{tabular}{|lrrrr|rrr|rrr|rrr|rrr|rrr|}
\hline
\multicolumn{5}{|l|}{\textbf{SUL}} &
\multicolumn{3}{c|}{\textbf{Resets (Learn)}} &
\multicolumn{3}{c|}{\textbf{Resets (Total)}} &
\multicolumn{3}{c|}{\textbf{CounterExs}} &
\multicolumn{3}{c|}{\textbf{WCT Learn [ms]}} &
\multicolumn{3}{c|}{\textbf{WCT Test [ms]}} \\
 & $|Q|$ & $|\Gamma|$ & $|X|$ & $|C|$ &
\multicolumn{1}{|c}{\SLstar} & \multicolumn{1}{c}{\SLlambda} & \multicolumn{1}{c|}{\SLCT} &
\multicolumn{1}{|c}{\SLstar} & \multicolumn{1}{c}{\SLlambda} & \multicolumn{1}{c|}{\SLCT} &
\multicolumn{1}{|c}{\SLstar} & \multicolumn{1}{c}{\SLlambda} & \multicolumn{1}{c|}{\SLCT} &
\multicolumn{1}{|c}{\SLstar} & \multicolumn{1}{c}{\SLlambda} & \multicolumn{1}{c|}{\SLCT} &
\multicolumn{1}{|c}{\SLstar} & \multicolumn{1}{c}{\SLlambda} & \multicolumn{1}{c|}{\SLCT} \\
\hline
channel-frame & 5  & 8  & 3 & 2 & \textbf{11} & \textbf{11}  & 15          & \textbf{24}  & 28           & 32     & \textbf{1} & 2          & 2  & 49 & 43 & \textbf{36} & 295 & \textbf{289} & 294 \\
abp-receiver3 & 6  & 10 & 3 & 2 & 489         & \textbf{88}  & 466         & 614          & \textbf{249} & 610    & 4          & 4          & 4  & 147 & \textbf{74} & 245 & \textbf{256} & 282 & 264 \\
palindrome    & 6  & 15 & 4 & 0 & 479         & \textbf{358} & 476         & 508          & \textbf{384} & 504    & 5          & 5          & 5  & 73 & 52 & \textbf{51} & 406 & 403 & \textbf{402} \\
login         & 12 & 19 & 4 & 0 & 436         & \textbf{244} & 433         & 509          & \textbf{300} & 512    & 3          & \textbf{2} & 3  & 86 & \textbf{54} & 67 & \textbf{301} & 303 & 310 \\
abp-output    & 30 & 50 & 1 & 2 & 363         & \textbf{208} & 311         & \textbf{590} & 4\,552       & 6\,151 & \textbf{5} & 11         & 11 & \textbf{142} & 151 & 696 & 260 & 175 & \textbf{154} \\
sip           & 30 & 72 & 2 & 0 & 487         & \textbf{233} & 345         & \textbf{934} & 3\,633       & 2\,772 & \textbf{9} & 15         & 16 & 370 & \textbf{347} & 353 & 194 & \textbf{149} & 160 \\ \hline
fifo3         & 12 & 16 & 4 & 0 & 29          & 24           & \textbf{23} & 212          & \textbf{202} & 209    & 5          & 5          & 5  & 114 & \textbf{106} & 108 & \textbf{547} & 636 & 563 \\
fifo5         & 18 & 24 & 6 & 0 & 66          & \textbf{55}  & 60          & 435          & \textbf{434} & 468    & \textbf{6} & 7          & 7  & 1\,303 & \textbf{1\,144} & 1\,451 & \textbf{575} & 600 & 584 \\
fifo7         & 24 & 32 & 8 & 0 & 118         & \textbf{96}  & 123         & \textbf{738} & 839          & 989    & \textbf{7} & 8          & 9  & 317\,435 & \textbf{279\,888} & 346\,897 & 589 & 591 & \textbf{583} \\
\hline
\end{tabular}
}
\end{table}

\medskip
\noindent
\textbf{Results.}
\Cref{tbl:wikiresults} summarizes the results of the
experiments in a black-box learning setup.
For every SUL, we report its complexity
(in number of locations~$|Q|$, transitions $|\Gamma|$, registers $|X|$,
and constants $|C|$) and, for each learning algorithm,
the number of resets (i.e., tests) during the learning phase,
total tests (incl.~counterexample search),
the number of counterexamples found,
and wall clock times (WCT) for learning and testing.
In \cref{tbl:wikiresults}, all numbers are averages from $20$ experiments.
It can be seen that the \SLlambda algorithm consistently outperforms the other
two algorithms w.r.t.~the number of tests during learning.
As can be expected, the \SLstar algorithm requires the fewest
counterexamples. Execution times do not show a consistent pattern for
these small systems or a clear `winner' between these three RA learning
algorithms, but there is a strong correlation between the number of learner
tests and the time that learning requires.
Due to this, in most cases, \SLlambda is fastest overall.

The SULs of the previous set of experiments were all quite small
($|\Gamma| \leq 72$), and did not show any scalability differences
between the three algorithms. Also, with the exception of fifo, the
benchmarks were not parametric.
In the following experiments, we scale the SULs which are learned.

\Cref{fig:dtls-mc} shows the results of our experiments with DTLS models.
For each algorithm, the graphs show the relationship between the number of
transitions in each hypothesis model and the number of resets with restricted
and unrestricted suffixes (in the first two graphs),
the number of counterexamples ($3^{rd}$ graph),
and execution times ($4^{th}$ graph).
It is evident that, with increasing model complexity, the number
of counterexamples grows linearly for all algorithms at roughly the same
rate, yet the number of resets grows much more rapidly for \SLstar
than it does for \SLCT and \SLlambda.
In terms of time performance, the trend is even more pronounced.
For SULs with more that $100$ transitions, learning times grow significantly
worse for \SLstar than the other two algorithms, and
\SLlambda clearly also beats \SLCT on even bigger systems.

\begin{figure}[t!]
\centering
\subfloat{
\includegraphics[width=.23\textwidth,height=8em]{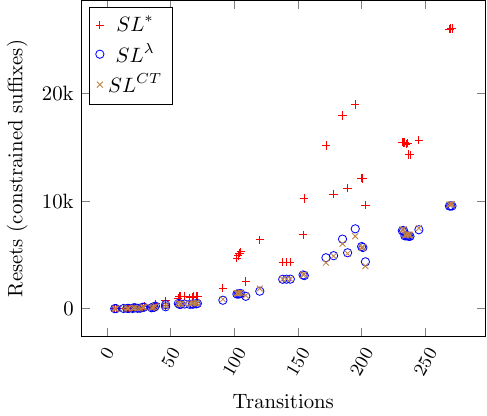}
}
\subfloat{
\includegraphics[width=.23\textwidth,height=8em]{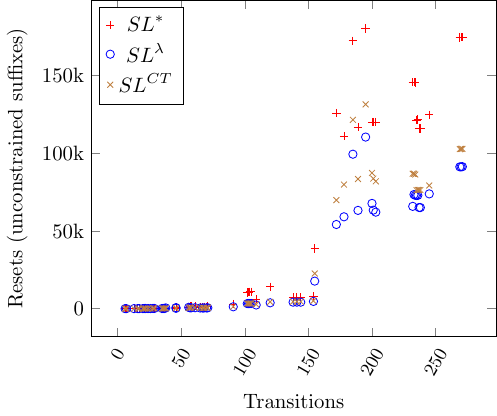}
}
\subfloat{
\includegraphics[width=.23\textwidth,height=8em]{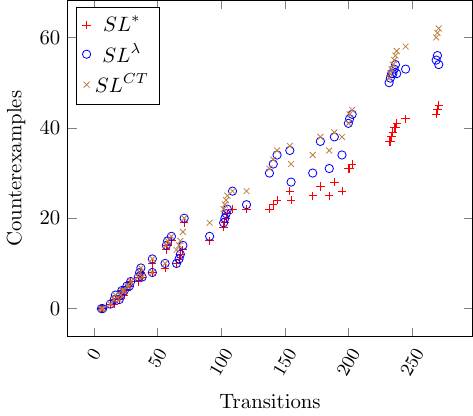}
}
\subfloat{
\includegraphics[width=.23\textwidth,height=8em]{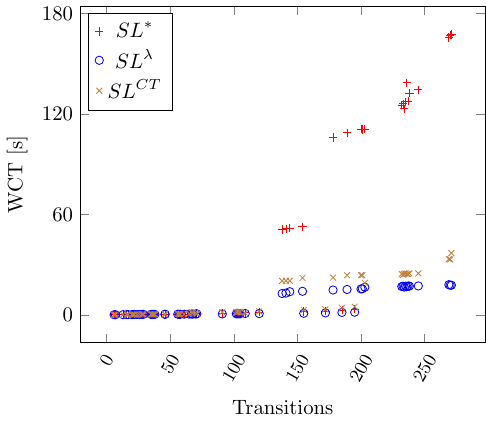}
}
\caption{Number of resets (two leftmost graphs), counterexamples ($3^{rd}$ graph),
and wall clock times ($4^{th}$ graph) for inferring models of the Mbed TLS
2.26.0 server.}
\label{fig:dtls-mc}
\end{figure}

\begin{figure}[t!]
\centering
\subfloat{
\includegraphics[width=.315\textwidth]{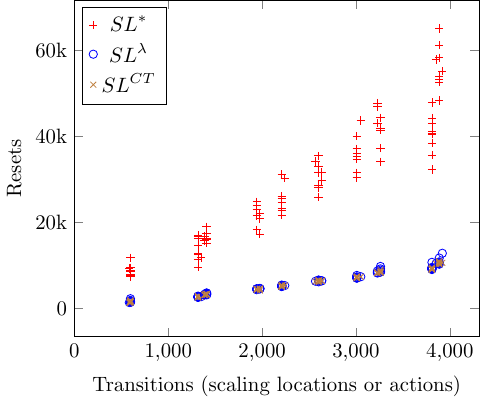}
}
\subfloat{
\includegraphics[width=.315\textwidth]{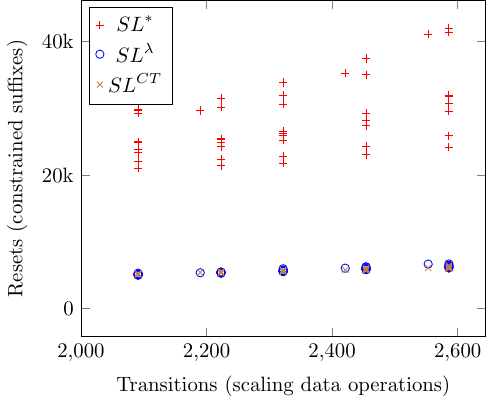}
}
\subfloat{
\includegraphics[width=.315\textwidth]{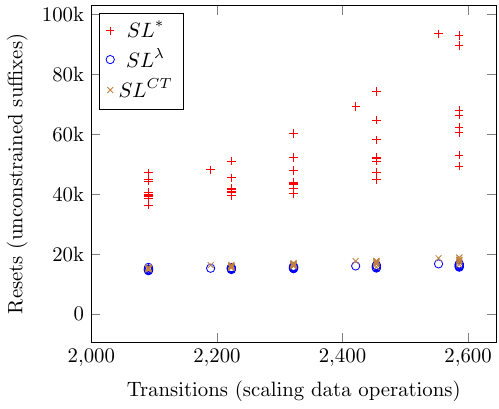}
}
\caption{Resets for inferring models of generated SULs, scaling the
number of transitions through locations and actions as well as by
increasing the percentage of transitions with data operations
using restricted and unrestricted suffixes.}
\label{fig:scaling-resets}
\end{figure}

Finally, \cref{fig:scaling-resets} shows the results of the experiments
with randomly generated automata. The graphs show how the number
of resets scales with the number of locations and actions (left) and
the number of registers when using both restricted (center)
and unrestricted suffixes (right).
The number of resets grows much more rapidly for \SLstar
than for the other algorithms. Not restricting suffixes leads
to a 2--4x increase in resets; notice the different scales on the y-axis.
Overall, the experiments show a clear advantage of \SLlambda over table-based
RA learning algorithms in terms of the number of resets and execution times
for bigger systems.
These results confirm the theoretical properties of the algorithms
and are consistent with the behavior of AAL algorithms for FSMs.

\section{Conclusion}
\label{sec:conclusion}

We have presented \SLlambda, a scalable tree-based algorithm for register automata learning.
\SLlambda reduces the membership queries needed for inferring RA models by constructing short
restricted suffixes incrementally.
This enables active learning in scenarios not feasible with previous algorithms.
We prove a reduction in the worst-case number of tests and, via a practical evaluation, show performance improvements on both real-world (i.e., on a complex network protocol) and synthetic models compared to the state-of-the-art RA learning~algorithm.

\subsubsection*{Acknowledgements}
This research was partially funded by
the Swedish Research Council (Vetenskapsr{\aa}det),
the Swedish Foundation for Strategic Research through project
\href{https://assist-project.github.io/}{aSSIsT},
the Knut and Alice Wallenberg Foundation through project UPDATE, and
the Deutsche Forschungsgemeinschaft (DFG, German Research Foundation) projects
\href{https://gepris.dfg.de/gepris/projekt/495857894}{495857894 (STING)} and
\href{https://gepris.dfg.de/gepris/projekt/442146713}{442146713 (NFDI4Ing)}.

\bibliographystyle{splncs04}
\bibliography{bibdatabase-utf8}

\ifwithappendix{
\appendix

\section{Example: Symmetries in an SDT}
\label{sec:symmetry}

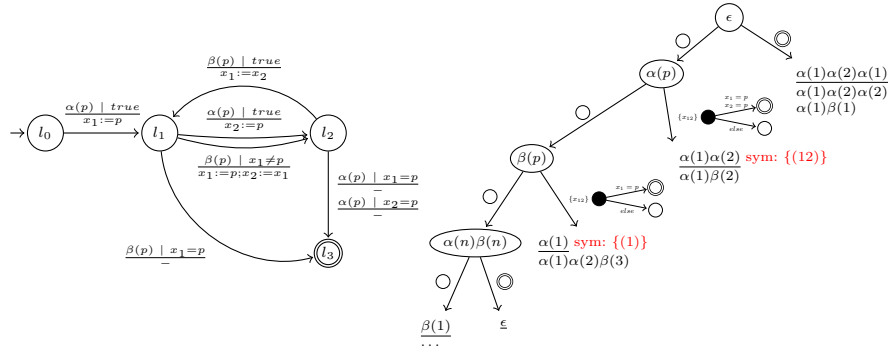
\begin{figure}[t]
  \begin{varwidth}{.39\textwidth}
    \centering
    \begin{tikzpicture}[shorten >=1pt,->,font=\scriptsize,scale=0.8,every node/.style={scale=0.8}]
  \tikzstyle{none}=[circle,minimum size=12pt,inner sep=0pt]
  \tikzstyle{final}=[circle, draw,style=double,minimum size=12pt,inner sep=0pt]
  \tikzstyle{vertex}=[circle,draw,minimum size=17pt,inner sep=0pt]
	
  \node[none]   (nn) at (0,0)  {};
  \node[vertex] (l0) at (0.8,0)    {$l_0$};
  \node[vertex] (l1) at (2.7,0)    {$l_1$};
  \node[vertex] (l2) at (5.5,0)    {$l_2$};   
  \node[final]  (l3) at (5.5,-2)   {$l_3$};   
  
  \draw (nn) -- (l0);
  
  \path (l0) edge[] node[above] {
    $\strans{\act(p)}{true}{x_1:=p}$} (l1);
  
  \path (l1) edge[bend right=5] node[above] {
    $\strans{\act(p)}{true}{x_2:=p}$} (l2);
  
  \path (l1) edge[bend right=50] node[below left] {
    $\strans{\beta(p)}{x_1=p}{-}$} (l3);
  
  \path (l1) edge[bend right=15] node[below] {
    $\strans{\beta(p)}{x_1 \neq p}{x_1:=p;x_2:=x_1}$} (l2);
  
  \path (l2) edge[] node[right, align=left] {
    $\strans{\act(p)}{x_1=p}{-}$\\
    $\strans{\act(p)}{x_2=p}{-}$} (l3);
  
  \path (l2) edge[bend right=50] node[above] {
    $\strans{\beta(p)}{true}{x_1:=x_2}$} (l1);
\end{tikzpicture}
  \end{varwidth}
  \hfill
  \begin{varwidth}{.59\textwidth}
    \centering
    \begin{lrbox}{\epsr}
  \begin{tikzpicture}[shorten >=1pt,->,font=\scriptsize,scale=0.5,every node/.style={scale=0.5}]
    \node[astate] {};
  \end{tikzpicture}
\end{lrbox}

\begin{lrbox}{\epsl}
  \begin{tikzpicture}[shorten >=1pt,->,font=\scriptsize,scale=0.5,every node/.style={scale=0.5}]
    \node[state] {};
  \end{tikzpicture}
\end{lrbox}

\begin{lrbox}{\xmid}
  \begin{tikzpicture}[shorten >=1pt,->,font=\scriptsize,scale=0.5,every node/.style={scale=0.5}]
    \node[fill=black,state,label=left:\{$x_{12}$\}] at (0,0) (l0) {};
    \node[astate] at (2,0.4) (l1) {};
    \node[state] at (2,-0.4) (l2) {};
    
    \path (l0) edge[] node [above,align=center] {$x_1=p$\\$x_2=p$} (l1);
    \path (l0) edge[] node [below] {$else$} (l2);
  \end{tikzpicture}
\end{lrbox}

\begin{lrbox}{\xleft}
  \begin{tikzpicture}[shorten >=1pt,->,font=\scriptsize,scale=0.5,every node/.style={scale=0.5}]
    \node[fill=black,state,label=left:\{$x_{12}$\}] at (0,0) (l0) {};
    \node[astate] at (2,0.4) (l1) {};
    \node[state] at (2,-0.4) (l2) {};
    
    \path (l0) edge[] node [above,align=center] {$x_1=p$} (l1);
    \path (l0) edge[] node [below] {$else$} (l2);
  \end{tikzpicture}
\end{lrbox}

\begin{tikzpicture}[shorten >=1pt,->,font=\scriptsize,scale=0.75,every node/.style={scale=0.75}]
  \node[below,dtnode] (eps) at (0,4.5) {$\epsilon$};
  
  \node[below,dtnode] (left)  at (-1.2,3.5) {$\act(p)$};
  \node[below,dtnode] (left2) at (-3.5,2)   {$\beta(p)$};	
  \node[below,dtnode] (left3) at (-4.5,.5)  {$\act(n)\beta(n)$};
  
  \node[below,align=left] (l0) at (-4,-1) {$\underline{\epsilon}$};
  \node[below,align=left] (l1) at (-5.2,-1) {
    $\underline{\beta(1)}$\\
	  \ldots
  };
  
  \node[below right,align=left] (l2) at (-3.5,.5) {
    $\underline{\alpha(1)}$ \textcolor{red}{sym: $\set{(1)}$} \\
    $\alpha(1)\alpha(2)\beta(3)$
  };
  
  \node[rectangle, below right,align=left] (l3) at (-1,2) {
    $\underline{\alpha(1)\alpha(2)}$ \textcolor{red}{sym: $\set{(12)}$}\\
	  $\alpha(1)\beta(2)$
  };
  
  \node[align=left,below] (l4) at (2,3.5) {
    $\underline{\alpha(1)\alpha(2)\alpha(1)}$\\
    $\alpha(1)\alpha(2)\alpha(2)$\\
    $\alpha(1)\beta(1)$
  };
  
  \path (eps)   edge[] node[right,yshift=.2em]  {\usebox\epsr} (l4);
  \path (eps)   edge[] node[left,yshift=.2em]   {\usebox\epsl} (left);
  \path (left)  edge[] node[left,yshift=.2em]   {\usebox\epsl} (left2);
  \path (left2) edge[] node[left,yshift=.2em]   {\usebox\epsl} (left3);
  \path (left3) edge[] node[left,yshift=.2em]   {\usebox\epsl} (l1);
  \path (left3) edge[] node[right,yshift=.2em]  {\usebox\epsr} (l0);
  
  \path (left)  edge[] node[right,yshift=0.1em] {\usebox\xmid} (l3.north west);
  \path (left2) edge[] node[right,yshift=0.1em] {\usebox\xleft} (l2);
\end{tikzpicture}
  \end{varwidth}
\caption{SUL (left) and Classification Tree (right) illustrating symmetry.}
\label{fig:symmetry}
\end{figure}

The classification tree in \cref{fig:symmetry} (right) represents 
all locations and transitions of the RA (left).
For the words in leaf $\alpha(1)\alpha(2)$ there are symmetries that could
lead to an incorrect remapping. The symmetry is not in the RA.
A check for register consistency finds information
to break it: in word $\alpha(1)\alpha(2)\beta(3)$ only $2$ is memorable.
This is a \emph{register inconsistency} and can be resolved by refining
the leaf of $\alpha(1)\alpha(2)$ by restricted suffix $\beta(n)\beta(p)$.

\section{Imposing Restrictions on Symbolic Suffixes}
\label{app:suffix-optimization}
As described in \cref{sec:ideas}, a tree query for a prefix $u$ and symbolic suffix $\symbsuff$ can be realized by a bounded number of membership queries for selected values of the data parameters of $\symbsuff$.
The number of these values can grow exponentially in the length of $\symbsuff$.
In order to reduce this number, we impose additional restrictions on the parameters of symbolic suffixes.
We consider two forms of restrictions on suffix parameters $p_i$:
\begin{inparaenum}[(i)]
	\item $\freshof{p_i}$, meaning that $p_i$ is different from all other preceding parameters in the prefix and suffix,
	\item $p_i = p_j$, where $j< i$, i.e., $p_j$ is an earlier parameter in the suffix.
\end{inparaenum}
How these restrictions are added to a symbolic suffix $\symbsuff$ differs depending on the context.
During analysis of counterexamples, symbolic suffixes are formed from concrete prefixes and suffixes, so a direct comparison can be made between parameters in the prefix and suffix.
When adding a symbolic suffix $\symbalpha\symbsuff$ to the classification tree, \eg, in a call to $\Refinefn(\prefixmap(u), \symbalpha \symbsuff)$, the parameters of $\symbsuff$ are symbolic, so comparisons between parameters in the prefix and suffix must instead be done using tree queries.
Let us explain how restrictions are added for these two different contexts.

\beginparagraph{Restricting Suffixes During Counterexample Analysis.}
During analysis of a counterexample $w = \act_1(\dval_1)\dots\act_n(\dval_n)$, we split $w$ into a prefix $\act_1(\dval_1)\dots\act_k(\dval_k)$ and a suffix $v=\act_{k+1}(\dval_{k+1})\dots\act_n(\dval_n)$.
When forming a symbolic suffix $\symbsuff$ from~$v$, restrictions on the parameters $p_i$ of $\symbsuff$ are obtained by
\begin{itemize}
	\item letting $p_i$ be fresh if $\dval_{k+i} \neq \dval_{j}$ for any $j < i$.
	\item letting $p_i$ equal a preceding parameter $p_j$ if $\dval_{k+i} = \dval_{k+j}$, and $p_j$ is fresh.
\end{itemize}
For example, suppose that we find a counterexample $\push(0)\push(1)\pop(1)\pop(0)$ for hypothesis $\hypo_0$ of \cref{fig:demo:stack:01}, and split it into prefix $u=\push(0)$ and suffix $\push(1)\pop(1)\pop(0)$.
When forming the symbolic suffix $\symbsuff=\push(p_1)\pop(p_2)\pop(p_3)$, we observe that $p_1$ can be fresh since the data value of $\push(1)$ does not equal any preceding data value.
Parameter $p_2$ can be restricted as equal to $p_1$ since the data value of $\pop(1)$ is equal to that of $\push(1)$, and $p_1$ is fresh.
The last parameter $p_3$ cannot be restricted since the data value of $\pop(0)$ is equal to a data value in the prefix, namely the data value of $\push(0)$.
With the restrictions $\langle fresh(p_1), p_2=p_1 \rangle$ on $\symbsuff$, only three membership queries must be made when performing the tree query $\treequerysul{u}{\symbsuff}$, as opposed to a total of fifteen membership queries that would be required if $\symbsuff$ were unrestricted.

\beginparagraph{Restricting Suffixes Added to the Classification Tree.}
Whenever a symbolic suffix is added to the $CT$, this suffix is formed by prepending a symbol $\symbalpha$ to a symbolic suffix $\symbsuff\in\symbsuffs$.
In this case, since the parameters of $\symbsuff$ are symbolic, we cannot compare them directly, so instead we compare guards from tree queries.

In the cases of register closedness or register consistency, the symbol $\symbalpha$ is formed from a concrete symbol $\dact{\act}{\dval}$ in a prefix $u\dact{\act}{\dval}$.
In these cases, $u$, $\symbalpha$ and $\symbsuff$ are chosen such that $\getmemorable{u\dact{\act}{\dval})}{\treeoracle}{\symbsuff}$ contains particular memorable parameters.
The parameters of $\symbalpha\symbsuff$ can be restricted by examining the guards of $\sdtfunction{u\dact{\act}{\dval}}{\symbsuff}$.
Let us denote the parameter of $\symbalpha$ as $p_1$ and the parameters of $\symbsuff$ as $p_2,\dots,p_{|\symbsuff|+1}$.
The restrictions on suffix $\symbalpha\symbsuff$ are then obtained by:
\begin{enumerate}
	\item
	letting the parameter of $\symbalpha$ be fresh if $\dval$ is not equal to any previous data value in $u$, and
	\item
	restricting each parameter $p_i$ with $i >1$ in $\symbalpha\symbsuff$ to be
	\begin{inparaenum}[(i)]
		\item fresh whenever $p_{i-1}$ is fresh in $\symbsuff$ or the branch taken in $\sdtfunction{u\dact{\act}{\dval}}{\symbsuff}$ for fresh $p_{i-1}$  reveals a sought register, and
		\item equal to a previous value $p_j$ in $\symbalpha\symbsuff$ if the branch taken in $\sdtfunction{u\dact{\act}{\dval}}{\symbsuff}$ for $p_{i-1}$ which is equal to the corresponding value reveals a sought register.
	\end{inparaenum}
\end{enumerate}
As an example, assume that we have a prefix $\act(0)\act(1)$ and symbolic suffix $\symbsuff = \act(p_2)\act(p_3)$.
We want to form an extended symbolic suffix $\symbalpha\symbsuff = \act(p_1)\act(p_2)\act(p_3)$, given the tree query $\sdtfunction{\act(0)\act(1)}{\symbsuff}$ shown in \cref{fig:suffopt-sdt-1}(left).
The registers $x_1$ and~$x_2$ are mapped to the data values 0 and 1, respectively, in the prefix.
In this example, the symbolic suffix $\symbalpha\symbsuff$ is needed to reveal the memorable data value $x_1$.
First, we restrict $p_1$ to be fresh, since the data value of $\act(1)$ does not equal any preceding parameter.
Second, we restrict $p_2$ to be equal to $p_1$, as the tree query has guard $p_2=x_2$, and $x_2$ corresponds to the fresh parameter $p_1$.
Finally, we note that $p_3$ cannot be restricted, as there is a guard $p_3=x_1$ and $x_1$ maps to a data value in $u$.

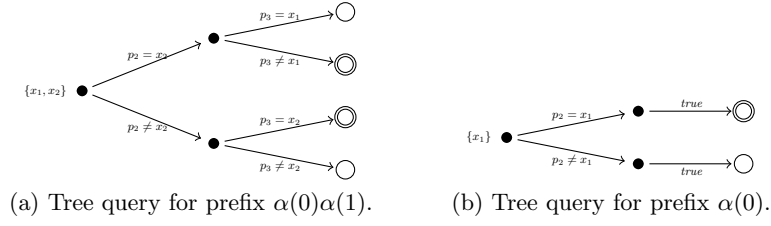
\begin{figure}[t]
  \centering
  \subfloat[Tree query for prefix $\act(0)\act(1)$.]{
    \begin{tikzpicture}[edgegraph,scale=1.75]
	\node[ustate,label={left:\{$x_1,x_2$\}}] at (0,0) (l0) {};
	\node[ustate] at (2,0.8) (l1) {};
	\node[ustate] at (2,-0.8) (l2) {};
	\node[state] at (4,1.2) (l3) {};
	\node[astate] at (4,0.4) (l4) {};
	\node[astate] at (4,-0.4) (l5) {};
	\node[state] at (4,-1.2) (l6) {};
	
	\path (l0) edge[] node [above] {$p_2 = x_2$} (l1);
	\path (l0) edge[] node [below] {$p_2 \neq x_2$} (l2);
	\path (l1) edge[] node [above] {$p_3 = x_1$} (l3);
	\path (l1) edge[] node [below] {$p_3 \neq x_1$} (l4);
	\path (l2) edge[] node [above] {$p_3 = x_2$} (l5);
	\path (l2) edge[] node [below] {$p_3 \neq x_2$} (l6);
\end{tikzpicture}
  }
  \hspace*{3em}
  \subfloat[Tree query for prefix $\act(0)$.]{
    \begin{tikzpicture}[edgegraph,scale=1.75]
	\node[ustate,label={left:\{$x_1$\}}] at (0,0) (l0) {};
	\node[ustate] at (2,0.4) (l1) {};
	\node[ustate] at (2,-0.4) (l2) {};
	\node[astate] at (3.6,0.4) (l3) {};
	\node[state] at (3.6,-0.4) (l4) {};
	
	\path (l0) edge[] node [above] {$p_2 = x_1$} (l1);
	\path (l0) edge[] node [below] {$p_2 \neq x_1$} (l2);
	\path (l1) edge[] node [above] {$\true$} (l3);
	\path (l2) edge[] node [below] {$\true$} (l4);
\end{tikzpicture}
  }
  \caption{Tree queries for two prefixes with symbolic suffix $\act(p_2)\act(p_3)$.}
  \label{fig:suffopt-sdt-1}
\end{figure}

The case of transition consistency (b), where two prefixes $u\dact{\act}{\dval}$ and $u\dact{\act}{\dval'}$ lead to the same location but are not equivalent under the identity mapping between registers, is handled similarly to register consistency.
Since $u\dact{\act}{\dval} \not\remapequiv{\treeoracle}{\symbsuff}{\mathbf{id}}  u\dact{\act}{\dval'}$, we need only consider one of $\sdtfunction{u\dact{\act}{\dval}}{\symbsuff}$ and $\sdtfunction{u\dact{\act}{\dval'}}{\symbsuff}$ when imposing restrictions on $\symbalpha\symbsuff$, with the caveat that one of $\dval$ and $\dval'$ will be equal to a parameter in $u$ so the parameter of $\symbalpha$ cannot be restricted.

In the case of location consistency, we want to find a symbolic suffix $\symbalpha\symbsuff$ that separates two prefixes $u$ and $u'$.
As such, when we restrict $\symbalpha\symbsuff$, we must do so in a way such that $\symbalpha\symbsuff$ retains its ability to separate $u$ and $u'$.
Assume that we have found two continuations $u\dact{\act}{\dval}$ and $u'\dact{\act}{\dval'}$ which lead to different locations.
Let us denote the parameter of $\symbalpha$ as $p_1$ and the parameters of $\symbsuff$ as $p_2,\cdots,p_{|\symbsuff|+1}$.
Now, let $\upath\in\domof{\sdtfunction{u\dact{\act}{\dval}}{\symbsuff}}$ be a $(u\dact{\act}{\dval},\symbsuff)$-path
and let $\upath'\in\domof{\sdtfunction{u'\dact{\act}{\dval'}}{\symbsuff}}$ be a $(u'\dact{\act}{\dval'}u,\symbsuff)$-path such that
\begin{inparaenum}[(i)]
	\item the conjunction of the guard expressions of $\upath$ and $\upath'$ is satisfied, and
	\item $\sdtfunction{u\dact{\act}{\dval}}{\symbsuff}(\upath) \neq \sdtfunction{u'\dact{\act}{\dval'}}{\symbsuff}(\upath')$.
\end{inparaenum}
For each such pair $\upath, \upath'$ of paths, a restricted symbolic suffix $\symbalpha\symbsuff_{(\upath,\upath')}$ is obtained by
\begin{enumerate}
	\item letting $p_1$ be fresh if $\dval$ does not equal any data value in $u$ and $\dval'$ does not equal any data value of $u'$,
	\item letting each parameter $p_i$ with $i>1$ be fresh if, in both $\upath$ and $\upath'$, the guard for $p_i$ is either $\true$ or a disequality guard, and
	\item restricting each parameter $p_i$ with $i>1$ as equal to a preceding parameter $p_j$ if
	\begin{inparaenum}[(i)]
		\item the only guard on $p_i$ in $\upath$ is $p_i=p_j$,
		\item the only guard on $p_i$ in $\upath'$ is $p_i=p_j$, $p_i\neq p_j$ or $\true$, and
		\item $p_j$ is fresh.
	\end{inparaenum}
\end{enumerate}
We choose as our restricted suffix $\symbalpha\symbsuff$ the $\symbalpha\symbsuff_{(\upath,\upath')}$ with the smallest number of unrestricted parameters.
The case of transition consistency (a) is handled similarly, with the only difference being that, in this case, $u=u'$.

As an example, assume we have two prefixes $u\dact{\act}{\dval} = \act(0)\act(1)$ and $u'\dact{\act}{\dval'} = \act(0)$, which lead to different locations with lowest common ancestor in $CT$ being $\symbsuff=\act(p_2)\act(p_3)$.
The SDTs representing $\sdtfunction{\act(0)\act(1)}{\symbsuff}$ and $\sdtfunction{\act(0)}{\symbsuff}$ are shown in \cref{fig:suffopt-sdt-1}.
Note that in both SDTs, $x_1$ is mapped to the parameter of $\act(0)$ and $x_2$ is mapped to the parameter of $\act(1)$.
There are two pairs of paths $(\upath,\upath')$ to choose from which satisfy both conditions (i) and (ii), namely
\begin{inparaenum}[(1)]
	\item $\upath = (p_2=x_2, p_3=x_1)$ and $\upath' = (p_2=x_2, p_3:\true)$, or
	\item $\upath = (p_2\neq x_2, p_3=x_2)$ and $\upath' = (p_2\neq x_2, p_3:\true)$.
\end{inparaenum}
We can restrict $p_1$ as fresh, since $\dval$ does not equal any parameter in $u$ and $\dval'$ does not equal any parameter in $u'$.
For (1), we can restrict $p_2=p_1$ as it is equal to the parameter corresponding to $p_1$ in both $\upath$ and $\upath'$ (the parameter of $\act(1)$ for $\upath$ and of $\act(0)$ for $\upath'$), and $p_1$ is fresh.
However, we cannot place any restriction on $p_3$, since, in $\upath$, $p_3$ is equal to a data value in $u$ (namely, $x_1$).
For (2), we can restrict $p_2$ to be fresh, since the guard is a disequality guard for both $\upath$ and $\upath'$, and we can restrict $p_3$ to be equal to $p_1$ since $p_3$ is equal to $x_2$ (\ie, equal to $p_1$) in $\upath$ and $p_3$ has a $\true$ guard in $\upath'$.
Since we have one unrestricted parameter in $\symbalpha\symbsuff$ for (1), but none for (2), we choose the set of restrictions of (2).
Thus, we restrict $\symbalpha\symbsuff$ by $\langle fresh(p_1), fresh(p_2), p_3=p_1 \rangle$.

\section{Correctness and Complexity Proofs}
\label{app:correctness}

In this appendix, we establish the correctness and complexity properties
of \SLlambda.

\medskip
\noindent
\textbf{Lemma} \ref{lemma:ce}.
\textit{
A counterexample always leads to
a new short prefix (Case 1 of \cref{alg:analyze-ce})
or new prefix (Case 2 of \cref{alg:analyze-ce}).
}
\begin{proof}
We know that at every index $i$, for $u\in As(w_{1:i-1})$, it holds that 
\begin{enumerate}
\item $\treequeryhyp{u\dact{\act}{\repr ug}}{\symbsuff} \equiv \treequeryhyp{u'}{\symbsuff}$ and that
\item $\treequeryhyp{u}{\symbalpha\symbsuff}$ has a $g$-guarded subtree
  $\treequeryhyp{u\dact{\act}{\repr ug}}{\symbsuff}$.
\end{enumerate}
We also know that $\treequerysul{u_m}{\epsilon} \equiv \treequeryhyp{u_m}{\epsilon}$
for $u_m \in As(w_{1:m})$ because $\epsilon$ is the symbolic suffix of the root
in the classification tree and determines if the location of $u_m$ is accepting 
or rejecting. On the other hand, $\treequerysul{\epsilon}{\symbsuff} \not\equiv \treequeryhyp{\epsilon}{\symbsuff}$ for $\symbsuff = w_{1:m}$ since $w$ is a counterexample.

As a consequence, at some index $i$ it must either be the case that 
$\treequerysul{u\dact{\act}{\repr ug}}{\symbsuff} \not\equiv \treequerysul{u'}{\symbsuff}$
or that $\treequerysul{u}{\symbalpha\symbsuff}$ has a $g'$-guarded subtree that is 
not present in $\treequeryhyp{u}{\symbalpha\symbsuff}$.
When analyzing a counterexample, we make $u\dact{\act}{\repr ug}$ a short 
prefix (a prefix, respectively). 
Either a refinement occurs immediately, or next time we arrive at
the same check $u\dact{\act}{\repr ug}$ will be a short prefix 
(prefix, respectively) and the condition will not be satisfied. 
The algorithm will continue with the next case or index of the 
counterexample and the arguments given above apply. 
\qed
\end{proof}

Let $m$ be the length of the longest counterexample, $t$ the number of
transitions, $r$ the maximal number of registers at any location, 
and $n$ the number of locations in the final model.

\medskip
\noindent
\textbf{Theorem} \ref{thm:complexity}.
\textit{
\SLlambda infers a RA for regular data language $\Lang$ with 
$O(t)$ equivalence queries and
$O(t^2 \, n^r + tmn \, m^m)$ membership queries for 
sifting words and analyzing counterexamples.
}
\begin{proof}
Every counterexample will lead to progress: the counterexample
will produce a new transition or a new short prefix and a corresponding
refinement is guaranteed to occur before the next equivalence query.  
The canonic acceptor has a finite number of locations and transitions.
By construction, the final model will accept~$\Lang$.

The classification tree will have at most $n+2r+2t = O(t)$ inner nodes 
(since $t \geq n > r$) created through refinement operations in 
\cref{alg:main} on any path to a leaf. 
Suffixes at these nodes tree have at most $r$ unrestricted parameters.
The size of $\prefixes$ is limited by $t+1$,
resulting in at most $O(t^2)$ tree queries
A tree query in the classification tree results in 
at most $O(n^r)$ membership queries as prefixes are of length $n$ or shorter. 
The number of counterexamples is limited by $t$. For the longest 
counterexample of length $m$, it can be necessary to compute 
$mn$ tree queries that each can require $O(m^m)$
membership queries in the worst case.
\qed
\end{proof}

This is an improvement over the worst case estimate 
of $O(tr)$ equivalence queries for \SLstar~\cite{CHJS:faoc-16}.
\SLlambda also improves the worst case for membership queries
for sifting to $O(t^2 \, n^r)$
from $O(t^2r \, n^r)$ for filling the table in \SLstar.

For analyzing counterexamples, \SLlambda replaces \SLstar's $O(trm \, m^m)$
worst case estimate with $O(tmn \, m^m )$, where $t$ dominates $n$ and $r$.

}

\end{document}